\documentclass[letterpaper,twocolumn,11pt,accepted=2024-11-07,floatfix]{quantumarticle} 
\pdfoutput=1
\usepackage{color}
\usepackage{graphicx}
\usepackage{mathrsfs,amsmath,amsxtra,amssymb,amsthm,amsfonts,subcaption,braket,comment}
\usepackage[utf8]{inputenc}
\usepackage[english]{babel}
\usepackage[numbers]{natbib}

\PassOptionsToPackage{hyphens}{url}\usepackage{hyperref}

\newtheorem{lemma}{Lemma}[section]

\newtheorem{definition}[lemma]{Definition}

\newcommand{\NN}{\mathbb{N}}

\newcommand{\obs}{\mathcal{O}}

\newcommand{\dfour}{ D${}^4$ }

\begin{document}
%\begin{titlingpage}
\title{Modeling Short-Range Microwave Networks to Scale Superconducting Quantum Computation}

\author{Nicholas LaRacuente}
\affiliation{Indiana University Bloomington, Bloomington, IN 47404, USA}
\orcid{0000-0003-0966-9272}
\author{Kaitlin N. Smith}
\affiliation{Northwestern University, Evanston, IL 60208, USA}
\author{Poolad Imany}
\affiliation{Icarus Quantum Inc., Boulder, CO 80302, USA}
\author{Kevin L. Silverman}
\affiliation{National Institute of Standards and Technology, Boulder, CO 80305, USA}
\author{Frederic T. Chong}
\affiliation{University of Chicago, Chicago, IL 60642, USA}

\begin{comment}
\author{Nicholas LaRacuente$^{1,*}$, Kaitlin N. Smith$^{2}$, \\ Poolad Imany$^{5}$, Kevin L. Silverman$^4$, Frederic T. Chong$^3$\\
{\small $^1$Indiana University Bloomington, Bloomington, IN 47404, USA}\\
{\small $^2$Northwestern University, Evanston, IL 60208, USA}\\
{\small $^3$University of Chicago, Chicago, IL 60642, USA}\\
{\small$^4$National Institute of Standards and Technology, Boulder, CO 80305, USA}\\
{\small$^5$Icarus Quantum Inc., Boulder, CO 80302, USA}\\
%{\small$^{\dagger}$ These authors contributed equally.}\\
{\small $^{*}$nlaracu@iu.edu}
%,{\small$^{\wedge}$kns@uchicago.edu}
}
\end{comment}

\maketitle

\begin{abstract}
A core challenge for superconducting quantum computers is to scale up the number of qubits in each processor without increasing noise or cross-talk. Distributed quantum computing across small qubit arrays, known as chiplets, can address these challenges in a scalable manner. We propose a chiplet architecture over microwave links with potential to exceed monolithic performance on near-term hardware. Our methods of modeling and evaluating the chiplet architecture bridge the physical and network layers in these processors. We find evidence that distributing computation across chiplets may reduce the overall error rates associated with moving data across the device, despite higher error figures for transfers across links. Preliminary analyses suggest that latency is not substantially impacted, and that at least some applications and architectures may avoid bottlenecks around chiplet boundaries. In the long-term, short-range networks may underlie quantum computers just as local area networks underlie classical datacenters and supercomputers today.
\end{abstract}

\section{Introduction}

Quantum computing is at the forefront of emerging technologies, owing to its promise to solve some algorithms dramatically faster than classical counterparts.
Quantum computers (QCs) evaluate quantum circuits in a manner similar to classical computers assessing classical circuits. However, quantum information's ability to leverage superposition, interference, and entanglement is anticipated to provide QCs significant advantage in tasks
ranging from security-relevant computing \cite{shor1999polynomial} and optimization \cite{farhi2014quantum} to simulation of quantum dynamics \cite{kandala2017hardware}. To harness the power of quantum computation, quantum bits, or qubits, are needed which are the core individual element for information representation and storage. Useful QCs require many qubits that are interconnected to each other in a low-noise and fault-tolerant manner. Many different architectures have been used to encode qubits, such as superconducting circuits (SCs), trapped ions, neutral atoms, quantum dots, and photons. \cite{nielsen2010quantum}. %, such as superconducting circuits (SCs), trapped ions \cite{bruzewicz2019trapped-ion, kielpinski2002architecture}, neutral atoms \cite{ebadi2021quantum}, quantum dots \cite{hendrickx2021four}, and photons \cite{zhong2020quantum,arrazola2021quantum}.
To date, SC quantum computers have been one of the most promising platforms \cite{devoret2004superconducting, wendin2017quantum, kjaergaard2020superconducting}.
%The record number of SC qubits so far is 127-qubit processors from IBM~\cite{127q-release}, and a 66-qubit processor from Hefei National Laboratory \cite{wu2021strong}.
Recent demonstrations of quantum advantage have shown that interference with just over 50 qubits can be harnessed for computational speedups~\cite{arute2019quantum,wu2021strong,zhong2020quantum}. Additionally, early demonstrations of logical qubits are emerging~\cite{google2023suppressing, acharya2024quantum}. However, more and cleaner qubits will be required to exploit quantum computing for practical applications.

%These qubits however, are not error-corrected and their noise figures are too high to perform a useful classical algorithm on them. 

%Quantum information processing (QIP) provides an exciting new avenue for solving intractable problems based on the principles of quantum mechanics. Specifically, quantum computation implemented with quantum computers (QCs) will evaluate quantum circuits in a manner similar to classical computers. However, quantum information's ability to leverage superposition, interference, and entanglement is anticipated to provide QCs significant advantage in factoring~\cite{shor1999polynomial}, database search~\cite{grover1996fast}, and solving linear systems when large, error-corrected devices are available for quantum programming. Even in the near-term with limited error correction, QC use for chemistry~\cite{kandala2017hardware} and optimization~\cite{moll2018quantum} applications have been proposed on modestly-sized machines. 

State of art QCs, sometimes called Noisy Intermediate Scale Quantum (NISQ) devices, are error prone yet reaching a thousand-qubit capacities~\cite{preskill2018quantum}. The qubits on these QCs are not error corrected and often have limited connectivity that requires special mapping routines to transform an algorithm into a technology-dependent form~\cite{smith2019quantum}. NISQ qubits are characterized by their sensitivity to interactions with their environment and each other, and the gates and measurements required for meaningful computation are implemented with imperfect, error-prone operations. To achieve the realm of usefulness for quantum devices, many more qubits, approximately one million or more, and lower noise figures are essential~\cite{sevilla2020forecasting}. Concerningly, as noted in \cite{sevilla2020forecasting}, ``quantum computer designers face a trade-off between trying to optimize for quantum computers with many physical qubits and quantum computers with very low gate error rate."

Potentially nonlinear scaling of physical resources with qubit number or circuit size may impede useful quantum advantages. Some forms of near-term quantum advantage rely on precise ratios of quantum to classical resources \cite{bravyi2020quantum, maslov2021quantum}. Even larger quantum-classical separations could be in jeopardy under an inverse relationship between qubit quality and system size. Many long-term approaches to quantum computing, such as fault tolerance, conventionally require per-qubit noise to remain below a fixed threshold for arbitrarily large systems. The need to keep error rates bounded and total resource costs close to linear motivates approaches to reduce unwanted coupling within a single quantum computer \cite{quantum_intranet}.

Scaling up the qubit number without increasing noise comes across many challenges: (1) SC qubits operate inside a dilution refrigerator (DR) around 10s of mK, requiring a coax cable feed-through each to address individual qubits. The number of coax lines that can go into a DR poses an important engineering challenge in the number of qubits that can exist and operate inside each DR. To overcome this challenge however, using optical fibers and photodetectors instead of coax cables and use of wavelength division multiplexing to address multiple qubits with each fiber feed-through has been proposed \cite{lecocq2021control}. (2) Placing many qubits in close proximity to each other can increase their connectivity, helpful for engineered qubit-qubit interactions. However, this comes at the cost of increased cross-talk which causes the qubits to decohere faster~\cite{chamberland2020topological}. (3) Fabricating many qubits on a single wafer reduces the yield of processors with high qubit performance since the increased likelihood of a fault of can drastically impact the average fidelity of the processor as a whole~\cite{smith2022scaling}. Furthermore, due to the large footprint of SC qubits, about 1 $mm^2$~\cite{chow2014implementing}, even a 12-inch wafer can only house around 100,000 qubits. (4) A bottleneck facing SC qubit QCs, especially with a monolithic approach, involves designing device packaging that support increased qubit numbers while preserving qubit operator fidelities and coherence~\cite{huang2021microwave}. As qubit capacity increases on-chip, the chip's enclosure must provide efficient thermalization as well as suppress losses caused by spurious package modes. In addition, the management of chip I/O must be considered. 

All of these challenges point towards the use of a chiplet architecture and distributed quantum computing \cite{brecht2016multilayer,quantum_intranet}. To move towards modular QC systems, technical hurdles associated with loss, fidelity, latency, and scalibility of the interconnect technology must be addressed~\cite{awschalom2021development}. Nonetheless, increasingly large devices \cite{wu2021strong, 127q-release} are more likely to be error-limited for algorithms requiring many cross-device qubit movements, while smaller devices may achieve low noise without exceeding the range of classical state vector simulation. A grand challenge of quantum computing is to achieve the best of both. A major advance would be an architecture achieving competitive fidelities today and the ability to add connected qubits with little impact on attributes of pre-existing qubits.

%Additionally, near-term QCs have limited connectivity between qubits because spacial and noise constraints inflicted by the physical quantum chip influence intra-chip routing. All of these factors contribute to the challenges that restrict scalability. Even if the aforementioned design challenges are overcome, QCs are speculated to not scale past thousands of qubits on a single chip due to engineering hurdles associated with preserving qubit quality during bulk production, expanding refrigeration technology, and maintaining precise quantum control within a densely-packed space~\cite{national2019quantum}. A possible solution to these difficult noise and engineering challenges involves exploring the possibility of quantum intranets, or locally-connected devices, in the near-term~\cite{quantum_intranet}.

Classical computing saw similar challenges associated with scaling the system-on-chip (SoC) model~\cite{demir2014galaxy}. Integrated circuit (IC) yield loss is a result of manufacturing defects caused randomly by environmental particulates or by systematic sources such as unavoidable variation or errors during fabrication~\cite{berglund2003trends}. Because a larger chip has a higher probability of containing a defect that renders the entire system unusable, fabricating fewer, large SoCs on a wafer results in lower yields than ICs with a smaller wafer footprint. This observation has led to the increase in popularity of chiplet-based architectures for improved cost and flexibility in system design and use~\cite{chiplet-blog}. Just as pitfalls were eventually observed in classical computing devices taking a monolithic approach, it is anticipated that continuing to pack qubits on single chip will push the capabilities of manufacturing, cooling, and control technologies to their limit. Thus, to develop QCs containing the millions of qubits required fault-tolerance, we are motivated to pursue a modular approach based on multiple cores working in sync to accomplish computational tasks in a distributed manner. 

Distributed quantum architectures have been proposed in ~\cite{monroe2014large,cuomo2020towards,rodrigo2020will}. Related works present techniques for targeting a network of quantum chips for unified computing tasks~\cite{baker2020time,ferrari2020compiler}. Major players in quantum computing have noted the potential of distribution for scaling quantum computers ~\cite{quantum_intranet,power-of-networked-cluster}. Our work here aims to develop link modeling methods that help guide the development of next-generation, modular quantum architectures. Our main contribution is to bridge the physical and network layers in evaluating if and how short-range, high-performance links could accelerate progress in quantum computing. Our primary focus is distinct from conventional studies of quantum communication or networking applications, examining what underpins these networks instead of assuming common protocols. By connecting the physical with network and application layers, we begin to concretely answer questions about when and how linked chiplets obtain advantages over monolithic architectures. Maybe surprisingly, today's quantum processors and links are probably near the point at which chiplets begin to show benefits.

Modeling quantum networks is complicated due its wide range of possible technologies, paradigms, uses, and aspects. In this study, we focus on microwave links between nearby, superconducting processors. Motivating this focus is spectacular progress in microwave link hardware. Microwave links already achieve transfer fidelities and latencies on scales not far from local processing. Such performance underpins the proposal to use quantum links for the low-level implementation of quantum computers. In this picture, fault tolerance and entanglement distillation are not assumed as network abstractions. Instead, a chiplet architecture at the physical layer may underlie these and other abstractions, accelerating the path to relevant thresholds.

\subsection{Main contributions \& where this work fits}
\label{subsect:contribution}
A key contribution of this work is to bridge the gap from understanding microwave links at the physical layer to distributed quantum computing applications. Rather than encapsulate network overheads as abstract cost functions, we analyze how noise and costs in both cross-chiplet links and local processing impact the eventual fidelity and feasibility of computations. Our models start and extrapolate from observations of real hardware and experimental physics models. We propose practical, concrete schemes to detect and reduce errors on near-term links and processors. Broad themes in potential applications are evaluated for feasibility on devices with limited connectivity.

This work does not attempt to provide a complete picture of exactly how quantum intranets will function, as this would involve much speculation about future developments. Instead, the analyses herein should be a springboard for co-design. Quantum software researchers may use these analyses to answer the question of whether a proposed algorithm can reasonably and should be distributed on near-term quantum hardware. For experimentalists, these results should motivate efforts to integrate cutting edge, SC processors with microwave links, even before local fault-tolerance or efficient microwave to optical transduction. With these advances combined, quantum architects can design sophisticated computing systems that stitch all elements of the QC stack together.

The structure of this paper is as follows: In Section \ref{sec:previous-work}, we review some related literature including analogous approaches to similar questions. Section \ref{sec:physicallayer} is about modeling quantum links and local gates at the physical layer, including a comparison of Hamiltonians between optical and microwave links, a review of hardware challenges for local processing, and quantum channel models of relevant operations. We present a quantum channel model that captures realistic noise on both microwave links and local SWAP operations, setting up later comparions. In Section \ref{sec:networklayer}, we compare monolithic with chiplet architectures at the network layer, incorporating physical information to derive expected fidelities of qubit transfers in addition to topological metrics. We also derive a detailed channel model that includes both links and local swaps, which we use to propose a simple error detection scheme for the regime of high-quality local processing with lossy links. Finally, we conclude with some open problems and areas for potential follow-up research.

\section{Background}
\label{sec:previous-work}
Front runners in the effort to scale quantum computers are increasingly distributed designs as a solution to scaling challenges ~\cite{quantum_intranet,power-of-networked-cluster}. In this section, a sample of particularly relevant work in the distributed quantum computing space is reviewed. Since this space is large, this Section serves to frame the focus of this work in the context of related but distinct efforts.

\subsection{Two major regimes in quantum networking}
\label{subsect:quantum-internet}
A frequently described idea that is not the main focus of this work is the quantum internet~\cite{pirandola2016physics,wehner2018quantum,cuomo2020towards}, a long-range union comprising of numerous networks between many quantum computers. Information transfer within the quantum internet relies on establishing shared entanglement between remote nodes. Quantum information travels great distances over optical links, assisted by quantum repeaters that mitigate signal loss. These systems will be sophisticated as traveling quantum information works in synergy with classical transmissions to implement teleportation-based communication~\cite{cacciapuoti2019quantum}. Network end-nodes will comprise of a variety of QCs, heterogeneous in terms of base technology, and quantum sensing nodes. Many simulation tools are emerging as an effort to study how resources and applications are allocated in quantum network architectures ~\cite{matsuo2019quantum, coopmans2021netsquid, leone2021qunet} - see \cite[Section 6]{wu2021sequence} or \cite{azuma_tools_2021} for a summary of related work. A major challenge to date involves developing high-efficiency and high-fidelity methods from matter qubits used primarily and processors and mobile qubits used in transfers. A prime example is connecting SC processors with mobile, photonic qubits ~\cite{mirhosseini2020superconducting,lauk2020perspectives}. Progress is being made in the multidiscipinary effort to network quantum devices ~\cite{chung2021illinois,wu2021illinois,van2020long,du2022quantum-capable} at large distances.

In contrast, the primary focus of this paper is the possibility for quantum int\textit{ra}nets \cite{quantum_intranet}: short-range, low-latency, quantum networks near the physical layer. A quantum intranet addresses fundamentally different questions from the quantum internet. Rather than provide future applications for quantum computers, quantum intranets may accelerate progress toward those quantum computers being a reality. Relating the quantum internet with quantum intranets are two primary considerations:
\begin{enumerate}
\item As noted in \cite[Technical Recommendation 1]{nqco2021coordinated}, ``only a handful of anticipated use cases [for long-range quantum networks] have been identified," so governments and other sectors ``must continue to invest in research on the potential advantages (and associated requirements) of quantum networks to justify future development." Small, efficient, high-performance test beds for quantum networks promote exploration, prototyping, and innovation. Ideas tested and invented on short-range networks may preview uses for long-range, long-term networks at relatively low cost.
\item Underlying the quantum internet will be many abstractions and assumed basic protocols. Common ``primitives" include teleportation, entanglement distillation, quantum storage, routing, real-time feedback, and sufficient local processing to support these operations without draining resources from computation. Just as today's supercomputers and data centers use local networks to combine parallel subunits, large quantum computers may consist of locally connected chiplets. The quantum internet might never replace these local intranets, just as today's classical Internet has not replaced local networks.
\end{enumerate}
While the quantum internet is an active focus of both scientific research and institutional attention, the near-term impact of quantum intranets might be frequently understated. This paper primarily focuses on the potential for intranets as a competitive quantum computing architecture, rather than as a small analog of the quantum internet.

Despite major progress in and even commercial applications of quantum networks, an effective bridge between quantum communication and computing remains a challenge. Conventional approaches rely on future protocols and abstractions, including reliable quantum memory, entanglement distillation, and encodings that achieve finite rates with vanishing error. In contrast, because of microwave links' high transfer fidelity and low latency, short-range networks built on this technology show promise to improve computation without relying on these underlying abstractions. Instead, short, high-quality links could help local quantum computers reach the scale and quality at which operations such as encoding and distillation will be more feasible.

\subsection{Ongoing work in short-range link hardware}
\label{subsect:prervious-work-experiments}
Experimental work has shown that quantum communication between devices is preliminary yet feasible. In the space of microwave links, the work of~\cite{kurpiers2018deterministic} shows that superconducting qubits and microwave links can couple at high efficiency, mediated by coupling resonators. The systems in~\cite{zhong2021deterministic} and~\cite{magnard2020microwave} demonstrate high-fidelity %(80-91\%) 
qubit state transfer and entanglement between two nodes separated by a meter within one DR and five meters between separated cryostats, respectively. Similarly,~\cite{zhou2023realizing} describes SWAP operations between four quantum modules, each containing a single qubit, with a microwave quantum state router. Ref.~\cite{niu2023low} demonstrates state transfer in a network of five, four-qubit processors connected by aluminum cables. This list of demonstrations, which is not all inclusive, are promising as a cold waveguide provides a reliable means to transfer microwave signals between superconducting qubits, perhaps enabling near-term, networked QC architectures.

In Section \ref{sec:physicallayer}, we describe the physics and characteristics of these links, comparing them with matter-optical transduction.

\subsection{Algorithms and applications}
\label{subsect:previous-work-algorithms}
%Still needs polish?
Recent investigations have studied resource requirements of algorithms transpiled to multiple quantum processors. Additionally, methods to understand the impact of network topology on the distribution of entanglement within quantum networks have been proposed using graph-based analysis~\cite{bapat2018unitary,eldredge2020entanglement}. Studies have also examined the tradeoffs of teleporting information vs. gates for various network topologies%quantum arithmetic algorithms
~\cite{meter2008arithmetic}. This research will eventually enable application-level compilers that optimize use distributed quantum software according to cost functions based on link use~\cite{ghodsollahee2021connectivity}. For example, in the area of protocols that route network traffic, the work in~\cite{haner2021distributed} presents Quantum MPI, inspired by classical messaging passing interface (MPI), to standardize qubit messaging and entanglement distribution between nodes for distributed and parallel quantum computing. Much has yet to be defined in the space of quantum network protocols due to ambiguity in how quantum networks will be physically realized.

At the application level, past work has explored extending the capabilities of near-term quantum hardware by employing classical hardware with techniques such as entanglement forging~\cite{eddins2021doubling}, circuit cutting followed by reconstructive post-processing~\cite{tang2021cutqc}, and translating tensor network algorithms to quantum circuits~\cite{barratt2020parallel}. These methods, however, are intended for use on individual QCs coupled with both classical resources and runtimes that scale with the size of the targeted, quantum application. Hence these techniques are more designed to bypass distributing quantum computation than to realize it.

Finally, limited communication among non-local qubits drives the search for algorithms that are well-suited for the distributed space. Variational quantum algorithms (VQAs) have emerged as a prime candidate algorithm class for distributed quantum computing. For example, circuit partitioning and cutting was applied toward distributing quantum variational optimization over multiple QCs linked with classical communication channels ~\cite{saleem2021quantum}. Additionally the work in ~\cite{diadamo2021distributed} contributes a methodology for decomposing the accelerated variational quantum eigensolver ($\alpha$-VQE) algorithms for ansatz initialization in a distributed setting where available resources that connect quantum nodes include classical control networks and entanglement networks. 

\section{Physical components} \label{sec:physicallayer}
% of a distributed network
\subsection{On-chip hardware}

SCs are popular quantum technology that are expected to be favorable to scaling. They employ circuit components that demonstrate ``atom-like'' behavior, and the qubits can be customized to enable specific operation regimes and properties. Here, we focus on transmon QCs used by IBM~\cite{jurcevic2021demonstration}, a company with plans to debut 1000+ qubit processors by 2025~\cite{IBM-roadmap-new}.

The performance of a QC cannot be measured in number of qubits alone. Features such as connectivity, gate fidelity, state preparation and measurement (SPAM) errors, and coherence time must also be considered. Because of the 2-D nature of superconducting devices, qubit-qubit connectivity has largely been limited to nearest-neighbor communication. An example of this type of topology is a grid where all neighbors communicate, but limitations associated with device control along with cross-talk between circuit elements has lead to the popularization of the ``heavy-hex'' layout~\cite{heavy-hex-blog} where each on-chip qubit is connected to either two or three of its neighbors. Although room for improvement is seemingly limited with respect to on-chip connectivity using 2-D layouts, gate fidelity, SPAM errors, and coherence times are expected to improve with time. The single-number metric of Quantum Volume (QV) takes all of the aformentioned QC device properties into consideration, making it a valuable tool to benchmark quantum hardware~\cite{cross2019validating}. 

IBM quantum devices demonstrate average single-qubit gate infidelity ranging from $10^{-3} $ to $10^{-4}$, and average two-qubit gate infidelity is about an order of magnitude worse at around $10^{-2} $ to $10^{-3}$~\cite{jurcevic2021demonstration,IBMQS}. These values result from randomized benchmarking~\cite{randBench,magesan2011scalable,magesan2012characterizing}. While it may not be as critical as fidelity metrics, gate execution time should also be noted as it influences the amount of computation a QC can perform within coherence windows. Circuit Layer Operations Per Second (CLOPS) is another benchmarking technique developed by IBM that takes operator execution time into consideration (as well as fidelity)~\cite{wack2021quality}. On IBM transmon devices, single-qubit gates implemented with a microwave DRAG pulse require 10s of ns, while two-qubit gates implemented as either a direct controlled-NOT (CX) gate or a cross-resonance gate (CR) have a duration requiring 100s of $ns$~\cite{jurcevic2021demonstration}. Because two-qubit operations dominate in terms of infidelity and duration, they are often emphasized more in system cost than single-qubit operations.

\subsection{Link hardware}
\label{sec:link-hardware}

To transfer quantum information, photons are ideal candidates  owing to traveling with the speed of light and long coherence times. For these photonic links, low-noise and fast  quantum transfer with high probability of success is required to connect quantum processors in a \textit{quantum} manner. In terms of speed, all the quantum processing has to take place before the qubits decohere, which is on the order of 100s of $\mu s$ for SC qubits. Quantum links are similar to two-qubit gates in that they also connect two qubits to each other, but from different and sometimes distant quantum processors. Therefore, we should compare the performance of links with local two-qubit gates for each processor. 

Two types of photonic links are used in quantum networks: optical and microwave. Optical links can operate with low thermal noise at room temperature due to their high frequency (around 200 THz), and are compatible with fiber-optical infrastructures and even satellite communications \cite{liao2017satellite}, therefore one can envision a large-scale network of quantum processors. However, bridging the five orders of magnitude energy, or frequency, gap between SC qubits and optical photons is an outstanding task. Electro-optical approaches have been used to overcome this challenge\cite{fan2018superconducting,shao2019microwave}, with efficiencies approaching 0.02\% \cite{fan2018superconducting}. To achieve higher efficiencies and lower noise figures with current technologies, intermediate quantum systems have been proposed, such as mechanical acoustic resonators \cite{higginbotham2018harnessing,mirhosseini2020superconducting,forsch2020microwave} or solid-state quantum emitters \cite{bartholomew2020chip,imany2022quantum}. As an example, a system with one intermediate mechanical step is considered, where the microwave and mechanical frequencies are the same \cite{mirhosseini2020superconducting}. The large frequency difference between the mechanical and optical modes however, can be mediated by a strong optical pump red-detuned from an optical cavity by the mechanical frequency. In such cases, the interaction Hamiltonian between the optical and mechanical mode can be linearized from $g_{om}a_0^{\dagger}a_0(b_m+b_m^{\dagger})$ to $g_{om}( b_ma_o^{\dagger}+b_m^{\dagger}a_o)$, with $b_m^{\dagger} (a_o^{\dagger})$ and $b_m (a_o)$ denoting the creation and annihilation operators for the mechanical (optical) mode, respectively, and $g_{om}$ the coupling rate between optics and mechanics. The overall interaction Hamiltonian is then simplified to

\begin{equation}
\hat{H}_{int}=g_{qm}(\sigma_{ge}b_m^{\dagger}+\sigma_{eg}b_m)+ g_{om}( b_ma_o^{\dagger}+b_m^{\dagger}a_o)
\end{equation}

\noindent with $g_{qm}$ representing the coupling rate between the SC qubit and the mechanical mode. Recently, converting quantum information between SC qubits and optical photons, or quantum transduction, has been demonstrated with an overall efficiency of $10^{-5}$ and an addition of 0.57 noise photon per every transduced quanta \cite{mirhosseini2020superconducting}.

Microwave links on the other hand, consist of photons with frequencies in the GHz range. Even though these links have to operate at mK temperatures to avoid thermal noise, the similarity between their frequency to that of SC qubits means the possibility for direct coupling between these two quantum systems through microwave resonators with high efficiency \cite{zhong2021deterministic}. The principle of state transfer between a SC qubit through a microwave link is the SC qubit emitting a photon into the link when it relaxes from the excited to ground state. The interaction Hamiltonian of this process has the form

\begin{equation}
\hat{H}_{int}=g_m(\sigma_{ge}a^{\dagger}+\sigma_{eg}a)
\end{equation}

\noindent where $\sigma_{ge}$ and $\sigma_{eg}$ are the operators swapping the qubit between the ground and excited states. $a^{\dagger}$ and $a$ are creation and annihilation operators for a microwave photon, respectively, in a standing-wave mode and $g_m$ is the coupling rate between the SC qubit and the link channel.  Recently, a 1 m-long microwave link between two, three-qubit SC quantum processors was demonstrated, with a fidelity of 0.91 and success probability of 0.88 \cite{zhong2021deterministic}. These links also operate at speeds on the order of 100 $ns$, comparable to that of two-qubit gates on chip. However, due to the low energy carried by each photon, these links have to operate at 10s of mK temperature to keep the thermal noise low, which makes long-distance microwave links out of reach. In another promising demonstration, a 5 m-long link between two DRs was shown to connect two quantum processors for quantum state transfer~\cite{magnard2020microwave}. This multi-meter microwave link experiment reported fidelity of 0.86. It is suggested that state transfer fidelity could reach as high as 0.96 as processes are refined~\cite{magnard2020microwave}.

In this article, we focus on NISQ devices, and therefore, microwave links which show a more near-term path for scalability due to their reasonable fidelity, success probability, and speed. We show that chiplet architectures with microwave links \cite{gold2021entanglement} can result in integration of small arrays of high-performance SC qubits to achieve an overall higher performance compared to a monolithic / larger quantum processor.   

Modular quantum systems will consist of compute clusters, chiplets, characterized by high coherence times, robust gate sets, and dense connectivity between qubits. These compute qubits will likely have rigid design constraints associated with the devices in which they are housed that will limit scaling. Transmission lines between clusters will ease the burdens associated with scaling up qubits.

\subsection{Modeling links \& local SWAP operations} \label{sub:chanmod}
\label{modeling-link-based-and-local-SWAP-operations}
\begin{figure}[t!]
\centering
\includegraphics[width=0.4\textwidth]{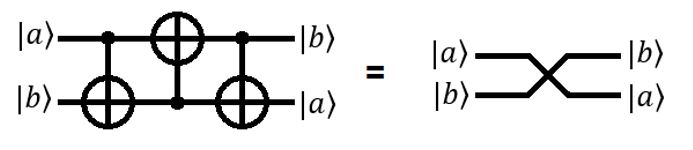}
\caption{CX implementation of SWAP operation between two qubits.}
\label{fig:swap-CX}
\end{figure}

A standard SWAP gate can be formed from 3 controlled-NOT (CX) gates (pictured in Figure~\ref{fig:swap-CX}). A quantum channel in finite dimension $d$ is fully characterized by its Choi matrix, the resulting density from the input of a $d \times d$ Bell pair through one side of the channel. We derive a rough estimate of the SWAP Choi matrix using channel tomography on qubit pairs within the \textit{ibmq\_brooklyn}, a 65-qubit processor. For this procedure, we applied Qiskit's built-in channel process tomography in parallel to neighboring qubits. For a two-qubit channel, the inferred Choi matrix has dimension $16 \times 16$. To simplify the notation we use brakets to express the computational basis states $\ket{0} ... \ket{15}$ and denote by $(\hat{1} / 16)$ the completely mixed state. Approximately, the inferred Choi matrix is
\begin{equation} \label{eq:duplexswapchoi} \scriptsize
\begin{split}
& 0.24 \ket{ \enspace 0}\bra{ \enspace 0} + 0.23 \ket{ \enspace 0}\bra{ \enspace 6} + 0.23 \ket{ \enspace 0}\bra{ \enspace 9} + 0.22 \ket{ \enspace 0}\bra{15} 
\\ + & 0.23 \ket{ \enspace 6}\bra{ \enspace 0} + 0.23 \ket{ \enspace 6}\bra{ \enspace 6} + 0.22 \ket{ \enspace 6}\bra{ \enspace 9} + 0.22 \ket{ \enspace 6}\bra{15}
\\ + & 0.23 \ket{ \enspace 9}\bra{ \enspace 0} + 0.22 \ket{ \enspace 9}\bra{ \enspace 6} + 0.23 \ket{ \enspace 9}\bra{ \enspace 9} + 0.22 \ket{ \enspace 9}\bra{15}
\\ + & 0.22 \ket{15}\bra{ \enspace 0} + 0.22 \ket{15}\bra{ \enspace 6} + 0.22 \ket{15}\bra{ \enspace 9} + 0.23 \ket{15}\bra{15}
\\ + & 0.07(\hat{1} / 16) .
\end{split}
\end{equation}

\noindent After calculating a Choi matrix as an average of 42 different single-swap Choi matrices between adjacent qubits, we average the result element-wise, apply basic readout error mitigation using a matrix pseudo-inverse, and drop elements with magnitude below $0.025$. After this process, the matrix is still not fully normalized, so the remaining contribution is assumed to be depolarizing. An ideal SWAP Choi matrix would be the density matrix corresponding to the state
\[ \frac{1}{2}(\ket{0000} + \ket{0110} + \ket{1001} + \ket{1111}) . \]
This estimate should be considered rough and likely underestimates the fidelity of a SWAP operation due to incomplete mitigation of readout and preparation errors.

We may alternatively regard a SWAP operation or link as a pair of single-qubit channels, moving a qubit state from the $i$th to $j$th physical qubit. Though this picture is less complete, it vastly simplifies the analysis of multiple moves as in Section \ref{sec:networklayer}, characterizing a single-qubit channel via tomography rather than a channel of dimension exponential in the length of the SWAP chain. Empirically, we observed in the experiments for Section \ref{sec:swchainchan} that a long chain model composed from single qubit SWAP tomographies more accurately recovers that of a long chain than does a reduced two-qubit model. We believe this discrepancy reflects cross-talk in nearby measurements or complications in readout error mitigation. The single qubit channel probably characterizes the SWAP more accurately by incorporating less readout noise. Here we again determine an average Choi matrix based on data from the 27-qubit $ibmq\_montreal$:
\begin{equation} \label{eq:oneswapchoi}
\begin{split}
\begin{pmatrix}
 0.4961 & & & 0.4726 \\
 & 0.0083 & & \\
 & & 0.0194 & \\
0.4726 & & & 0.4762
\end{pmatrix} .
\end{split}
\end{equation}
This matrix compares with the ideal given by $(\ket{00} + \ket{11})/2$.

For the microwave link, we obtain a rough Choi matrix of
\begin{equation} \label{eq:linkchoi}
\begin{pmatrix}
0.50 & & & 0.39 \\
& & & & \\
 & & 0.16 & \\
0.39 & & & 0.34 \\
\end{pmatrix}
\end{equation}
as an average of Choi matrices inferred from the plots in \cite{kurpiers2018deterministic} and \cite{magnard2020microwave}, and having fidelity of 0.97 between either of these and \cite{leung2019deterministic}. The transfer fidelity is roughly 0.81. Microwave link noise is almost completely amplitude damping. More recently, \cite{zhong2021deterministic} obtain a transfer fidelity of roughly 0.91. Though we do not infer a full Choi matrix from this newer result, we primarily use it in later analyses, inferring an amplitude damping channel with $\eta \approx 0.12$.

The Choi matrix captures only the noise aspect of the link and local SWAPs, omitting latency. Though links are often presumed slow, latency does not appear to be a major problem for microwave links. Our analysis of the above papers finds a typical link latency of roughly 200 $ns$. A CX gate on the \textit{ibmq\_montreal} takes roughly 400 $ns$, leading to a 1200 $ns$ local SWAP time. Even with microwave-optical transduction, the latency of a link is likely to be on the scale of microseconds, only slightly worse than local qubit movements. 

The time required to reset a link between uses is often noted as a potential bottleneck, but it does not appear to be a fundamental constraint. The 50 kHz rate from \cite{kurpiers2018deterministic} suggests a waiting time of around 20 $\mu s$ before link re-use, and \cite{magnard2020microwave} may have a time as slow as 300 $\mu s$. These slow rates arise, however, from issues of tuning that in principle would be improved by orders of magnitude in an experiment optimized for repetition. The links used in \cite{zhong2021deterministic} reset on timescales closer to their latency, hundreds of nanoseconds. Furthermore, each transfer on this system is a two-way exchange, doubling the effective bandwidth in many scenarios. In the future, link repetition times will plausibly be as short as $1-4\mu s$.

In contrast to a common assumption that links are slow compared to local processing - it seems that local links will achieve similar timescales as on-chip SWAPs. Rather than focus on latency, we therefore shift to issues of network topology and transferred state fidelity.

% Robust link technology is required for quantum chiplet systems to become a reality. Transduction technology that would allow room-temperature optical communication between QCs is still in its early stages. Fortunately, super-cooled microwave links have recently shown promise during experimental demonstrations of state transfer and entanglement~\cite{zhong2021deterministic,magnard2020microwave.}.

\section{Main Results: modeling distributed architectures \& comparing to monolithic}  \label{sec:networklayer}
There are many ways to define the local connectivity of a QC architecture, where popular candidates include two-dimensional grid-like connectivity and IBM's heavy-hex layout, representing different tradeoffs between connectivity and noise \cite{nation2021ibm}. Some previous analyses have evaluated alternative architectures based on the costs of entanglement creation \cite{eldredge2020entanglement,bapat2018unitary}. Sophisticated connectivity graphs such as trees and expanders may show advantages over straightfowardly two-dimensional layouts. We start by examining several layouts that appear to be strong or popular candidates.
\subsection{Connectivities and graph metrics}
\label{sub:graph-diameter}
\begin{figure*}[!tb] \scriptsize %\centering
	
	\begin{subfigure}[b]{0.32\textwidth} \centering
		\includegraphics[width=0.80\textwidth]{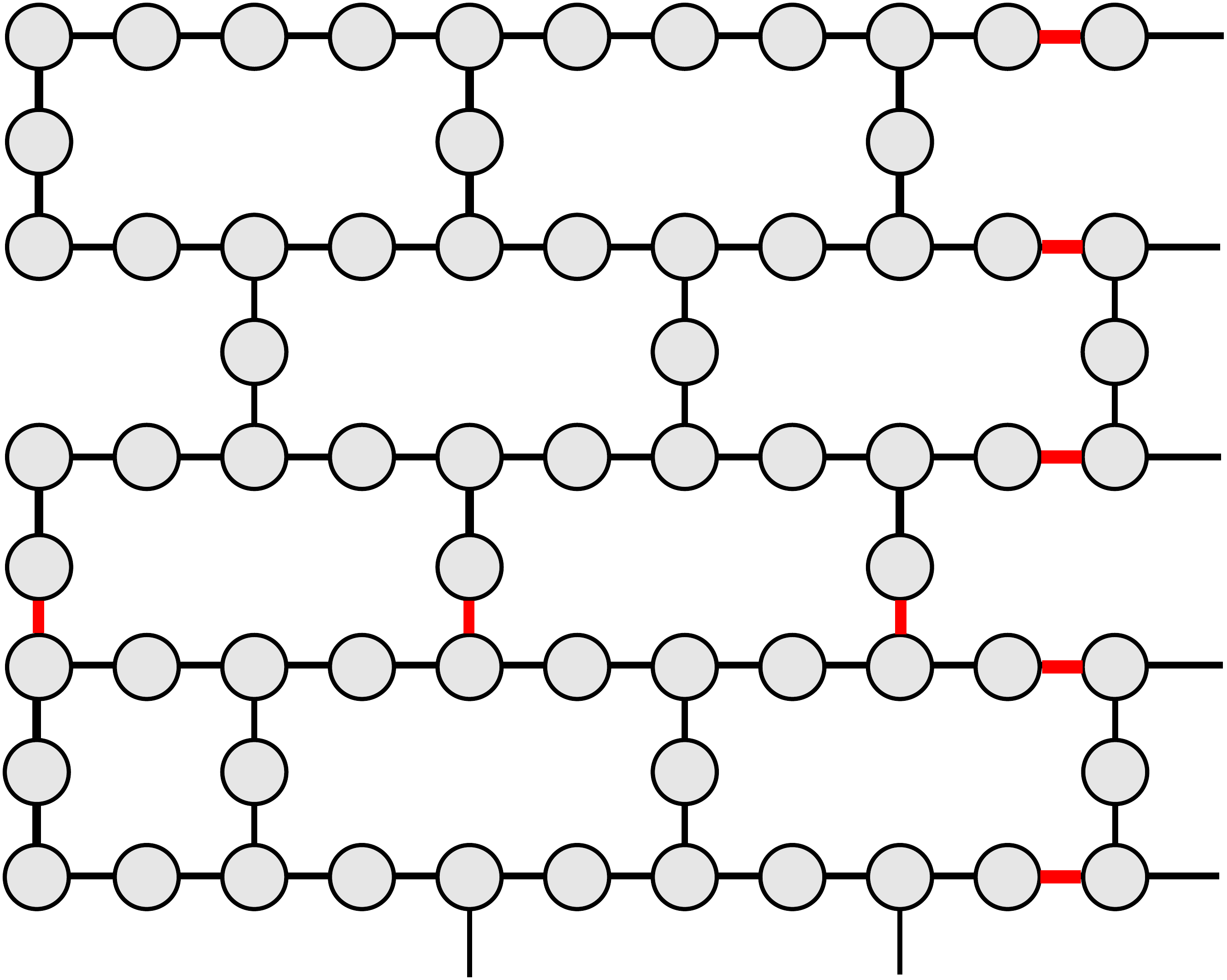}
		\caption{}
	\end{subfigure}
	\begin{subfigure}[b]{0.42\textwidth} \centering
		\includegraphics[width=0.98\textwidth]{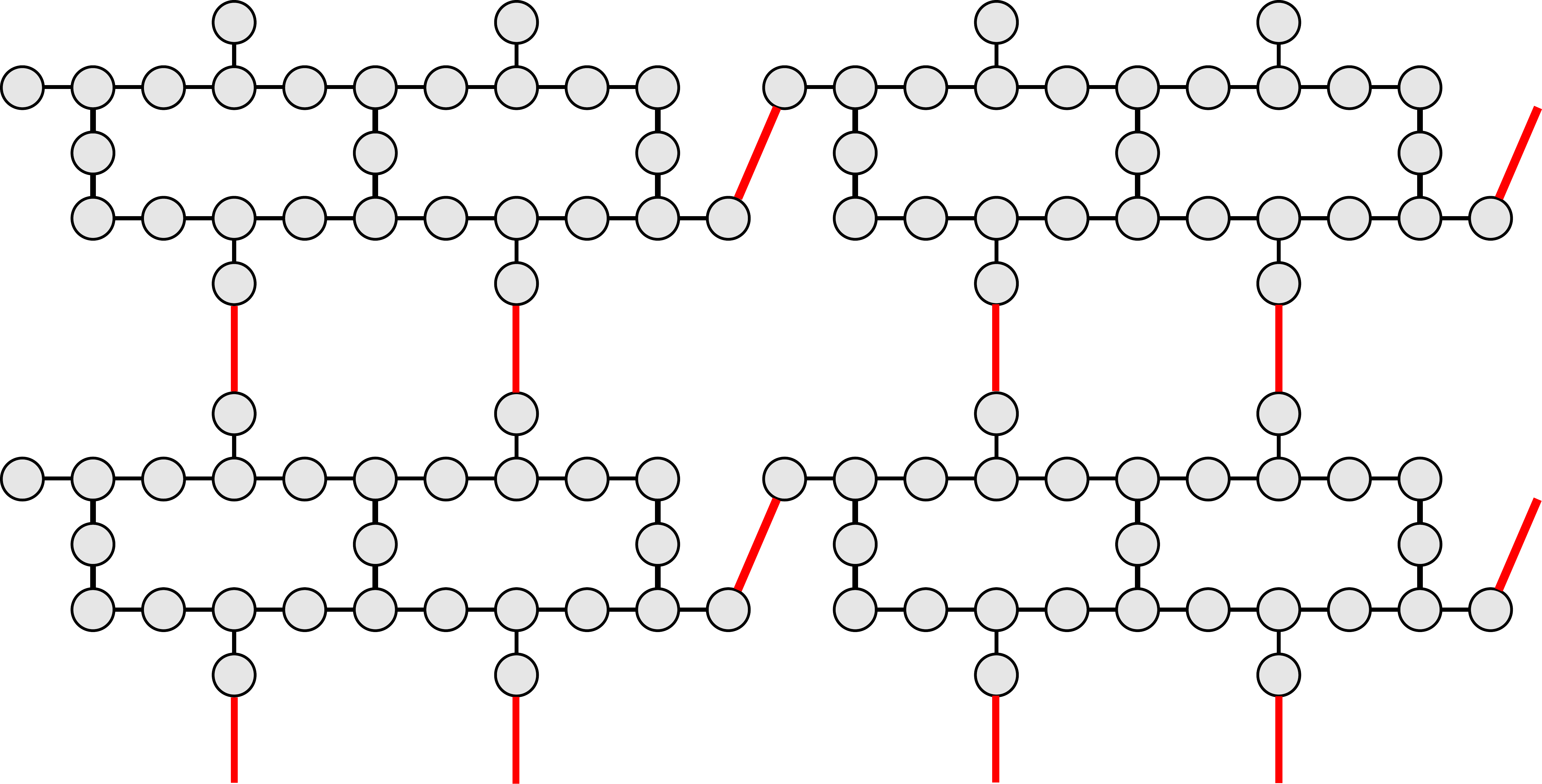}
		\vspace{2mm}
		\caption{}
	\end{subfigure}
	\begin{subfigure}[b]{0.32\textwidth} \centering
		\includegraphics[width=0.50\textwidth]{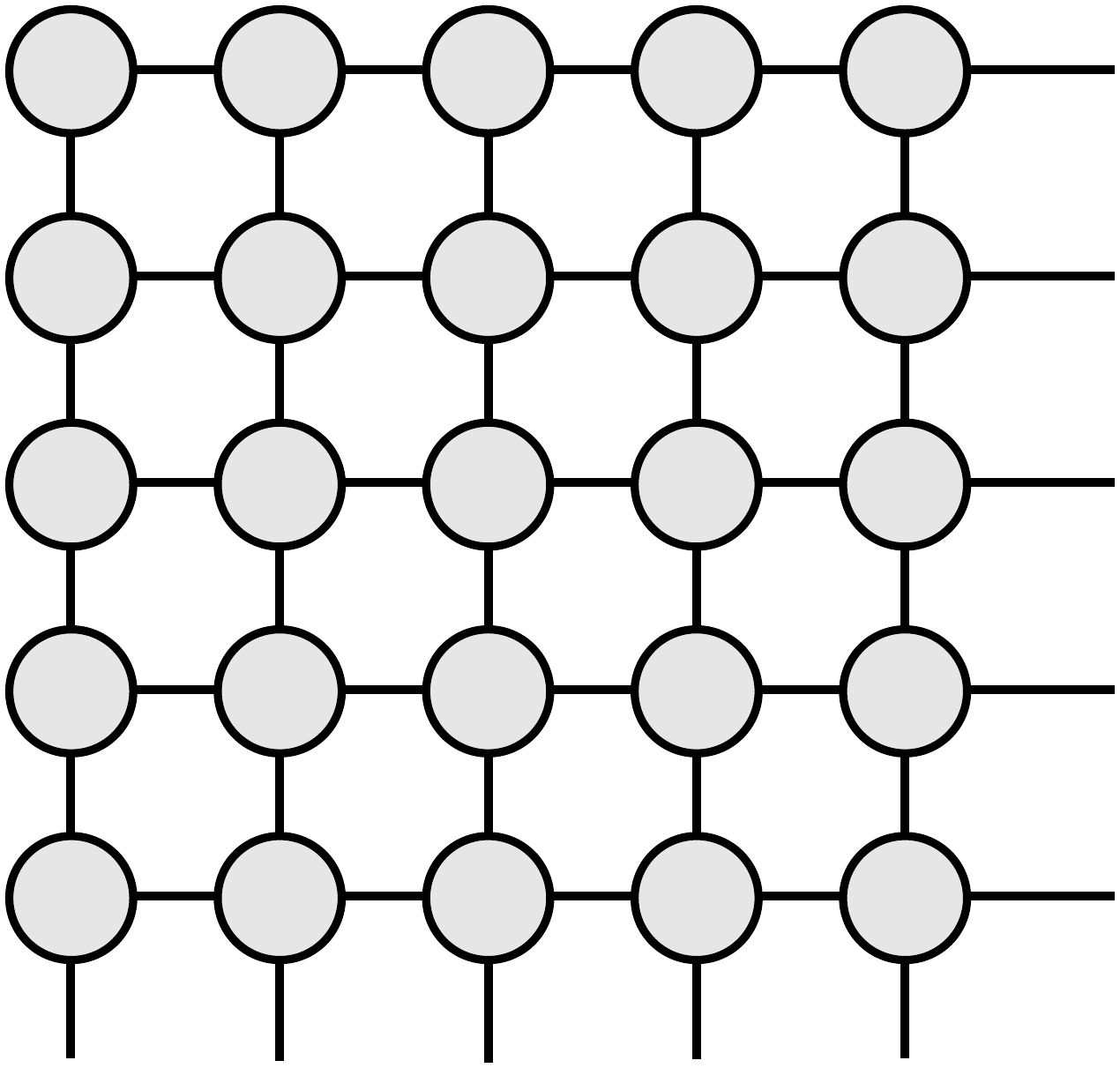}
		\vspace{8mm}
		\caption{}
	\end{subfigure}
	\begin{subfigure}[b]{0.32\textwidth} \centering
		\includegraphics[width=0.80\textwidth]{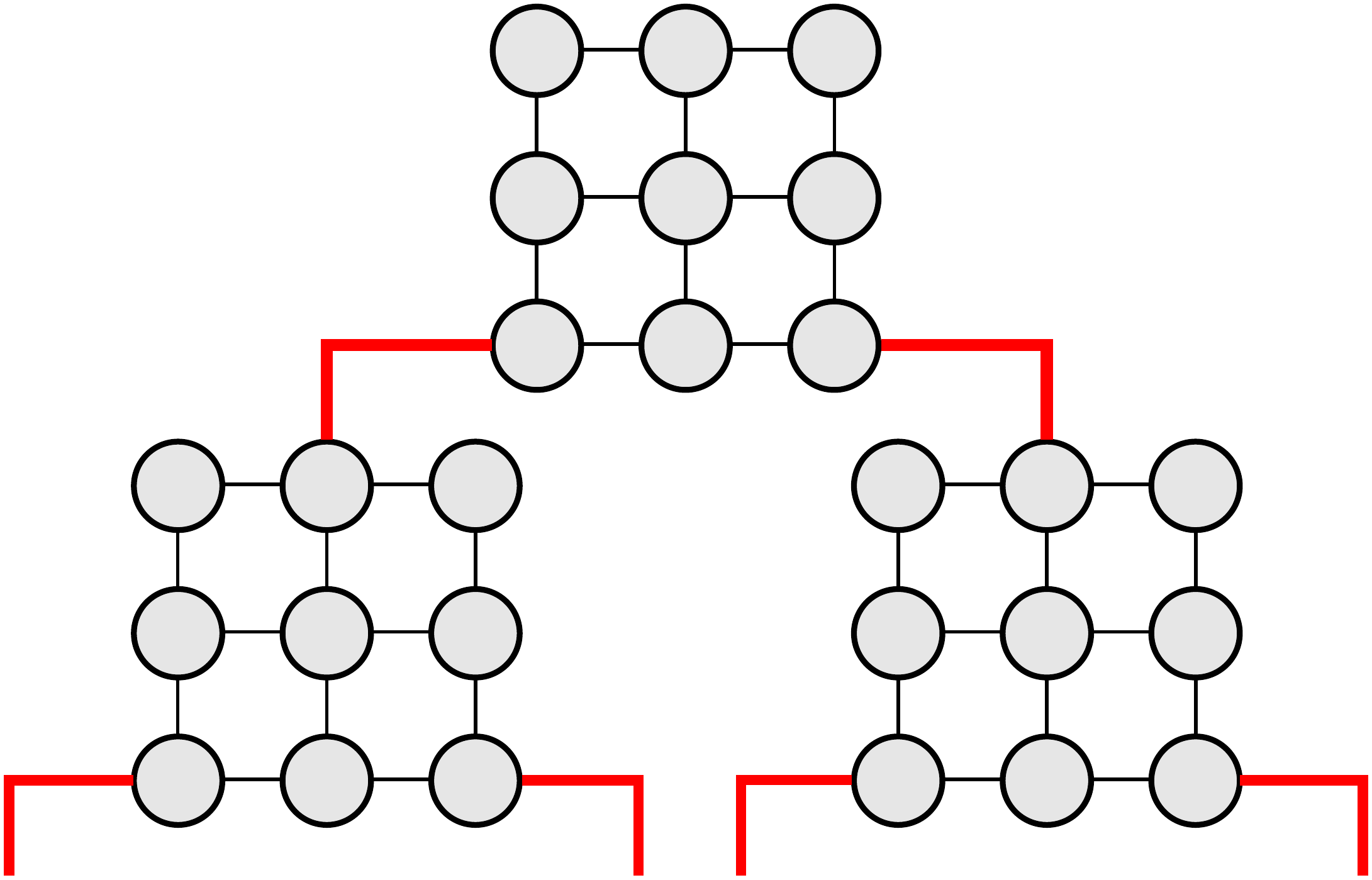}
		\vspace{8mm}
		\caption{}
	\end{subfigure}
	\begin{subfigure}[b]{0.32\textwidth} \centering
		\includegraphics[width=0.80\textwidth]{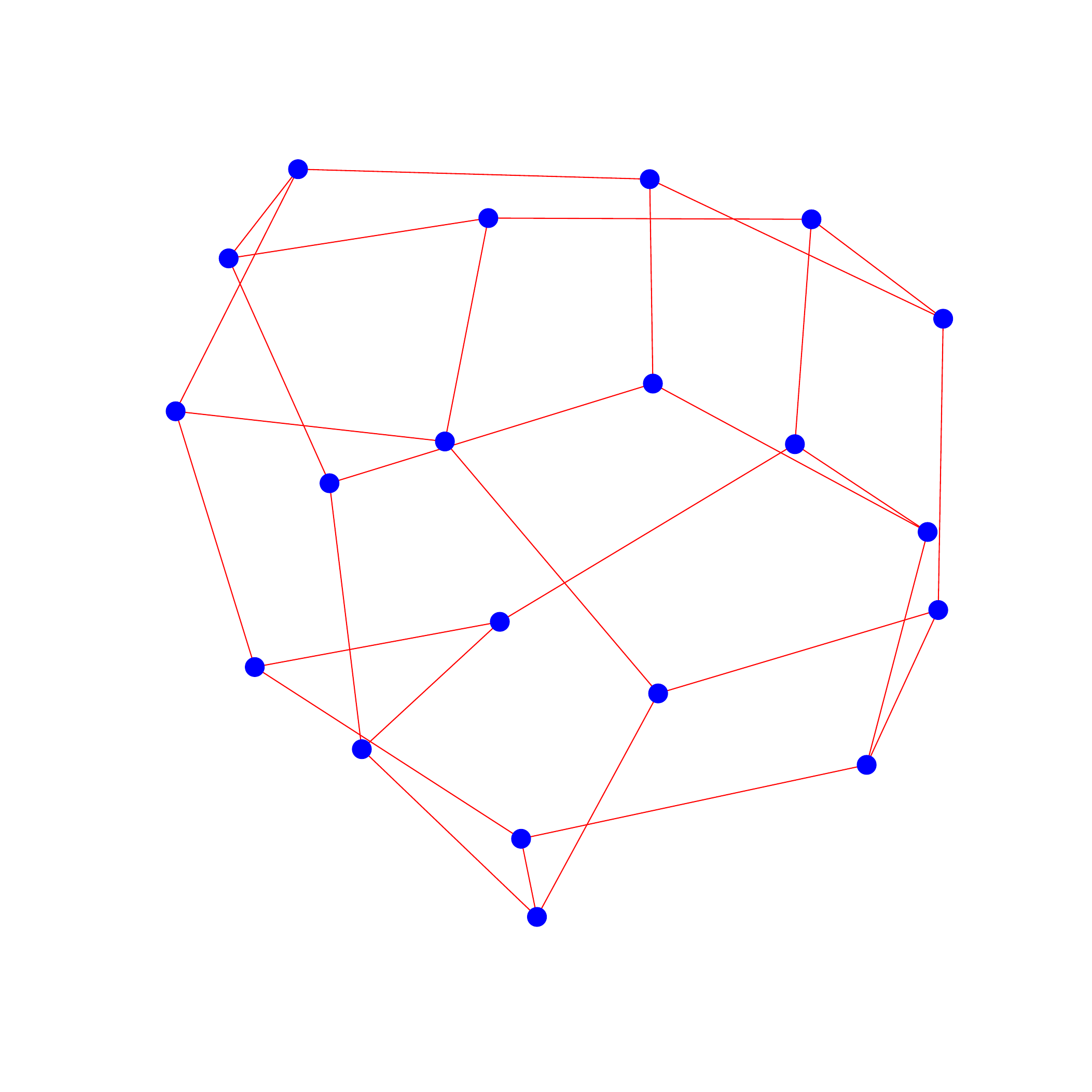}
		\caption{}
	\end{subfigure}
	\begin{subfigure}[b]{0.32\textwidth} \centering
		\includegraphics[width=0.80\textwidth]{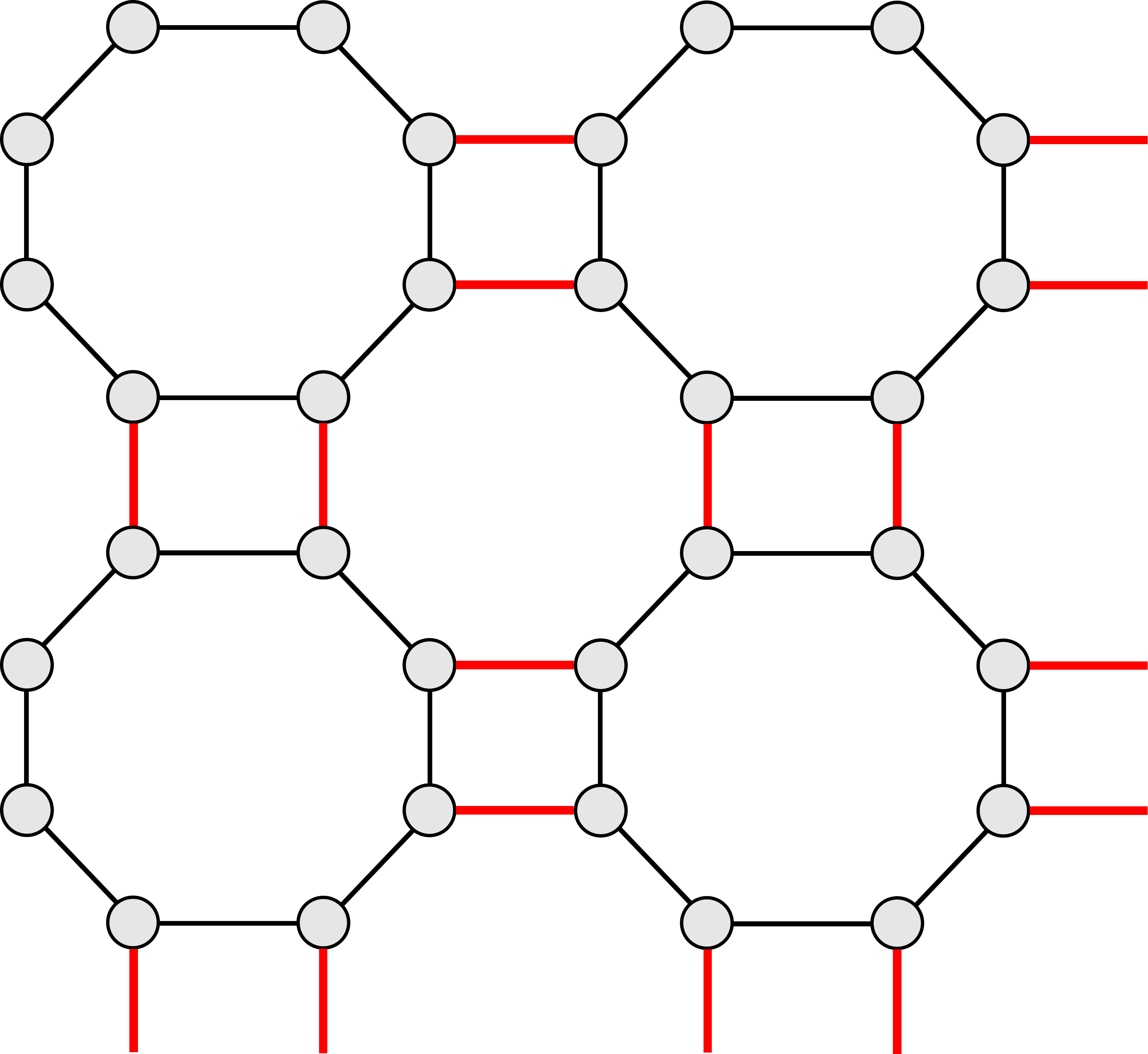}
		\caption{}
	\end{subfigure}
	\begin{subfigure}[b]{0.32\textwidth} \centering
		\includegraphics[width=0.80\textwidth]{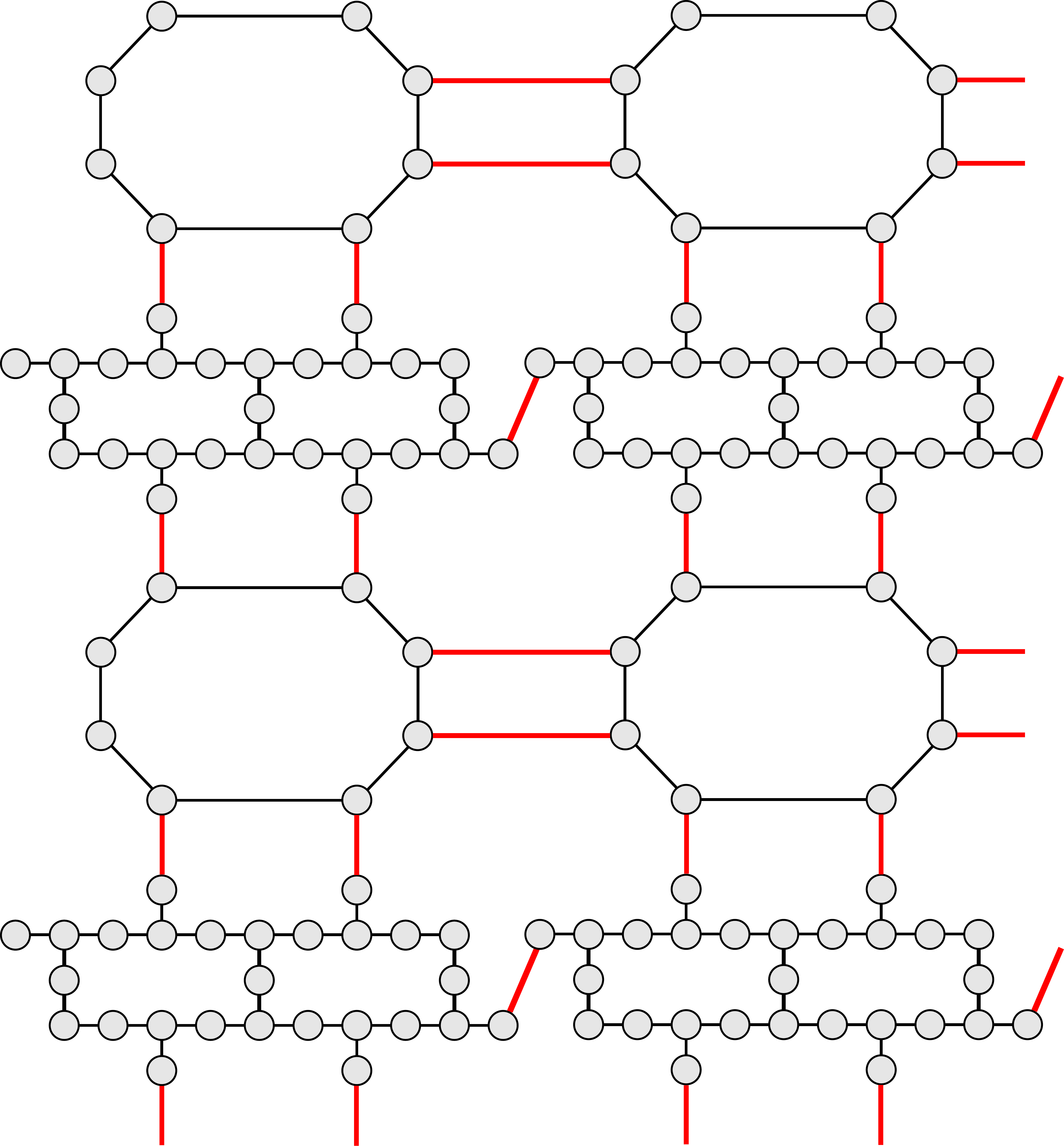}
		\caption{}
	\end{subfigure}
	\begin{subfigure}[b]{0.32\textwidth} \centering
		\includegraphics[width=0.80\textwidth]{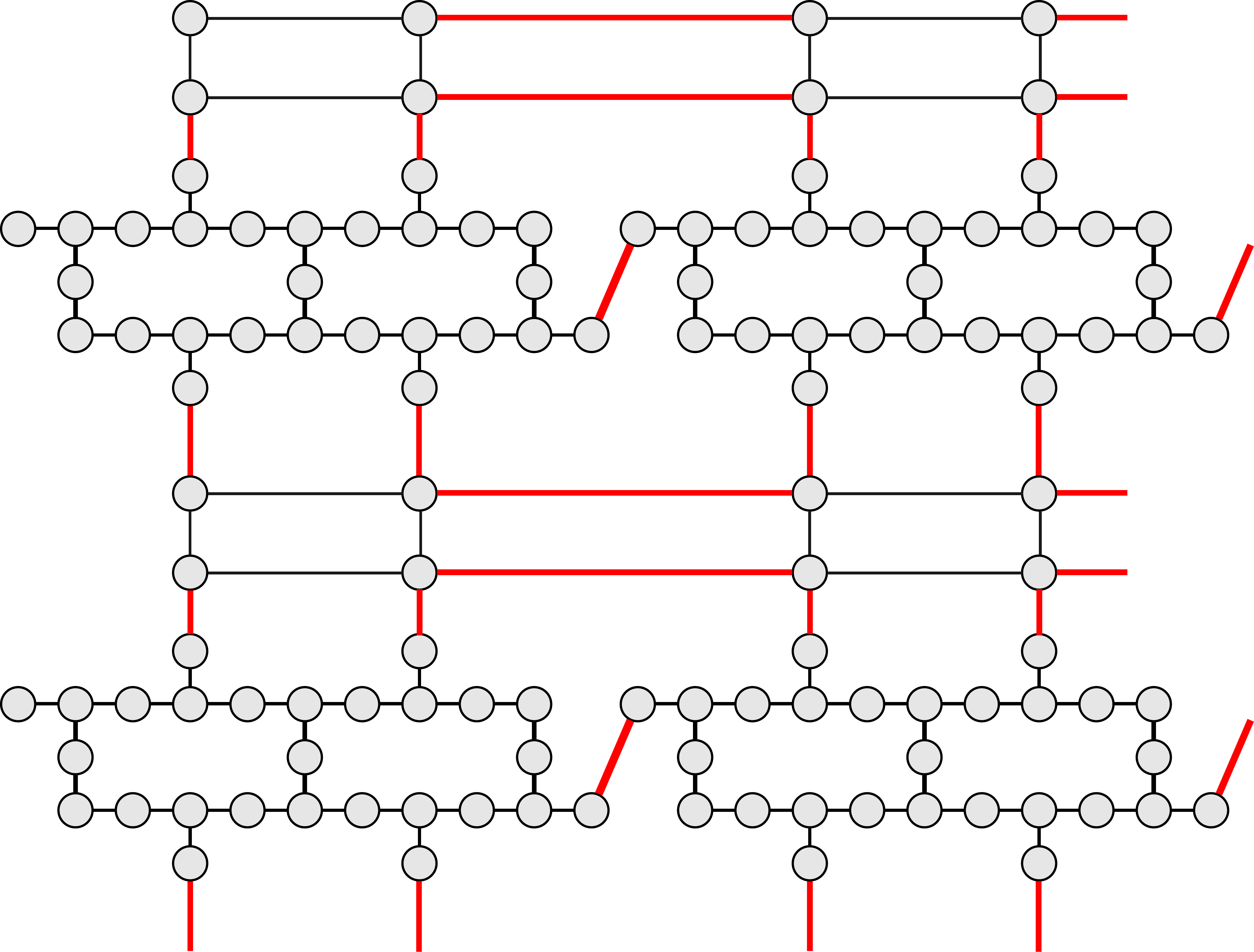}
		\caption{}
	\end{subfigure}
	\caption{The black lines connecting physical qubits indicate on-chip connections while the red lines indicate inter-chip microwave links. (a) Heavy-hex, where red lines illustrate possible link placements in 38-q chiplet. (b) Four 27-q chiplets with links on nodes of local coordination number 1. (c) Monolithic grid. (d) Tree of $3 \times 3$ grids. (e) Random graph with degree 3 and good spectral gap. Here each node (shown in blue) would be replaced by connections to a chiplet. (f) Octagonal chips are connected via links. (g) Octagonal chips are stacked with heavy-hex in vertically alternating layers. (h) Octagonal chips are stacked with squares in vertically alternating layers. \label{fig:topology-compare}}
\end{figure*}

We consider a variety of graph topologies. For those involving links, we note the expected ratio of on-chiplet qubits to links. Since each link would generally involve two chiplets, we calculate this number by dividing the number of qubits on a chiplet by half the number of links entering/leaving that chiplet. When chiplets are assembled into a distributed chip, this number roughly equals the ratio of total qubits to total links. There might be slightly fewer links in total if edges of the full chip do not have links, since there is nothing to link to.
\begin{enumerate}
	\item As a monolithic comparison point, we consider the standard two-dimensional grid. Each qubit not on the boundary has four neighbors.
	\item Removing some links from the monolithic grid yields a chiplet grid containing the same qubit number but distributed in 25-qubit blocks. We use this as a chiplet grid architecture. Since each local subgrid has up to 4 links, the qubit-to-link ratio is at least 12.5.
	\item Inspired by the study of \cite{bapat2018unitary}, we consider a tree of 25-qubit grids. Each non-root node connects to its parent via the middle qubit on top, and each non-leaf connects to each of its children through a bottom corner. With 3 links per chiplet, the qubit-to-link ratio is at least $16.6$.
	\item As a second monolithic comparison point, we consider a monolithic heavy-hex chip. The overall structure is chosen to be as square as possible for a given chip number.
	\item We split the heavy-hex layout into chiplets by replacing some connections with links. This has little effect on topology but may change the error tradeoffs as described in Subsection \ref{sub:est-fid}. In particular, we partition the chiplet into 80-qubit blocks, each with up to 16 links for a qubit-to-link ratio of at least 10. A smaller version of this layout is shown in Figure \ref{fig:topology-compare}(c). Unlike the grid case, here we can maintain a reasonable qubit-to-link ratio while keeping the heavy-hex structure across the entire device consistent. We expect topological properties similar to the monolithic heavy-hex but impose a rectangular overall structure, which might be less square than its monolithic counterpart due to the minimum block size.
	\item We consider linking 27-qubit IBM Falcon processors in a layout that nearly preserves heavy-hex structure, as illustrated in Figure \ref{fig:topology-compare}(b). There are 6 qubits of a Falcon processor that each have local coordination number one. Since these 6 qubits are relatively isolated, they may intuitively serve as natural joining points, buffering the main processors from potential noise around links. This layout yields a qubit-to-link ratio of at least 9.
	\item Depending on the link medium, it may sometimes be possible to bypass the constraints of two-dimensional layouts. Expander graphs may achieve high connectivity with relatively few links, avoiding bottlenecks by allowing links to criss-cross in a more complex structure. Each of these connects to such a point on another Falcon processor for qubit-to-link ratio of 9. To generate expander graphs, we use a well-known result that random graphs frequently have good expansion properties~\cite{puder2015expansion}. For each desired chip number, we search 800 random expander graphs and 60 random ways of embedding chiplets within each for that with the largest spectral gap, a figure of merit described below.
    \item We model a grid layout built on octagons. Octagon-based chips were considered in some Rigetti architectures \cite{noauthor_rigetti_nodate}.
    \item We model a hybrid layout in which octagon chiplets alternate with 27-qubit heavy-hex chiplets. The different chip geometries are stacked vertically. Here the octagon chips are arranged so that each qubit has two on-chip connections and one microwave link. The 27-qubit chiplets have links only attached to qubits with coordination number 1, such that each has 23 qubits that are not directly involved in an inter-chip link. The qubit-to-link ratio is 5.
    \item We model a hybrid layout between square and 27-qubit heavy-hex chiplets. Each qubit in the square has two links and two on-chip connections. Again, the 27-qubit chiplets each have 23 chiplets that are not directly involved in an inter-chip link, and only qubits with coordination number 1 are involved in links. The qubit-to-link ratio is about 4.4.
\end{enumerate}

Though some figures of merit could be derived analytically, others are somewhat complicated, and the random generation of expander graphs is best automated. Hence the analyses in this Subsection are computed using the NetworkX library, a standard tool for graph analysis~\cite{hagberg2008exploring}.

%\onecolumngrid
%\twocolumngrid

\begin{figure}[h!]
\includegraphics[width=0.49\textwidth]{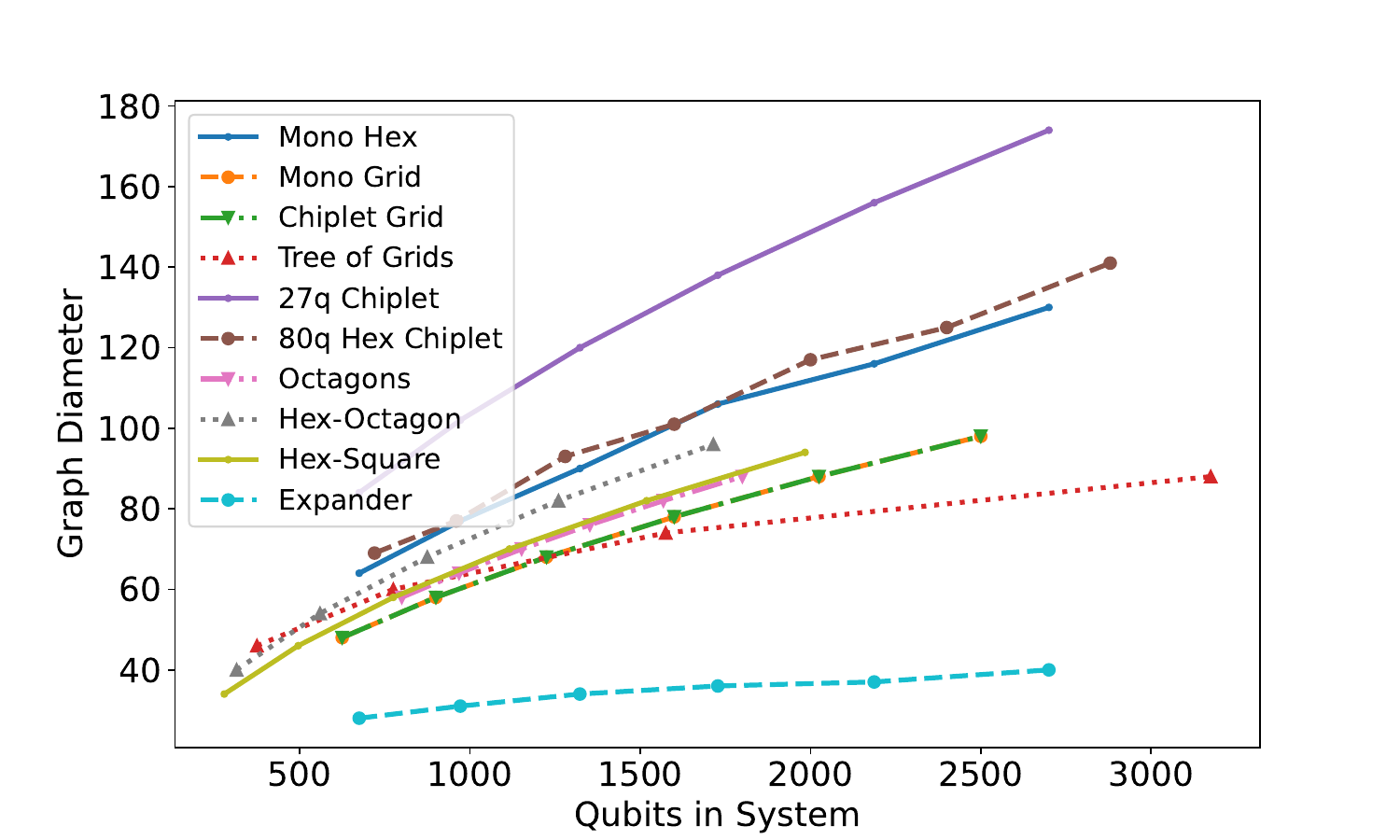}
\caption{Comparison of QC topology graph diameter vs. number of qubits for proposed architectures. Graph diameter quantifies the maximum number of swap operations needed for a qubit to transit from one end to another.}
\label{fig:graph-diam}
\end{figure}

The maximum time required to move a qubit via SWAP operations, including link uses, is proportional to the graph diameter of the system's connectivity graph.
\begin{definition}
    The graph diameter is given by taking maximum over all pairs of nodes of the shortest path between those nodes.
\end{definition}
Figure \ref{fig:graph-diam} shows graph diameters for the above topologies. In some cases, the graph diameter is not hard to calculate by hand. For the example of a square grid, the graph diameter is equal to twice the side length minus one, and it is exactly the same for the chiplet grid architecture. For heavy hex, one may estimate the graph diameter by counting the number of connections needed to move from left to right, then from top to bottom, and again add up effective side lengths. Graph diameter for a tree or expander is logarithmic in node number.

\begin{figure}[h!]
\includegraphics[width=0.49\textwidth]{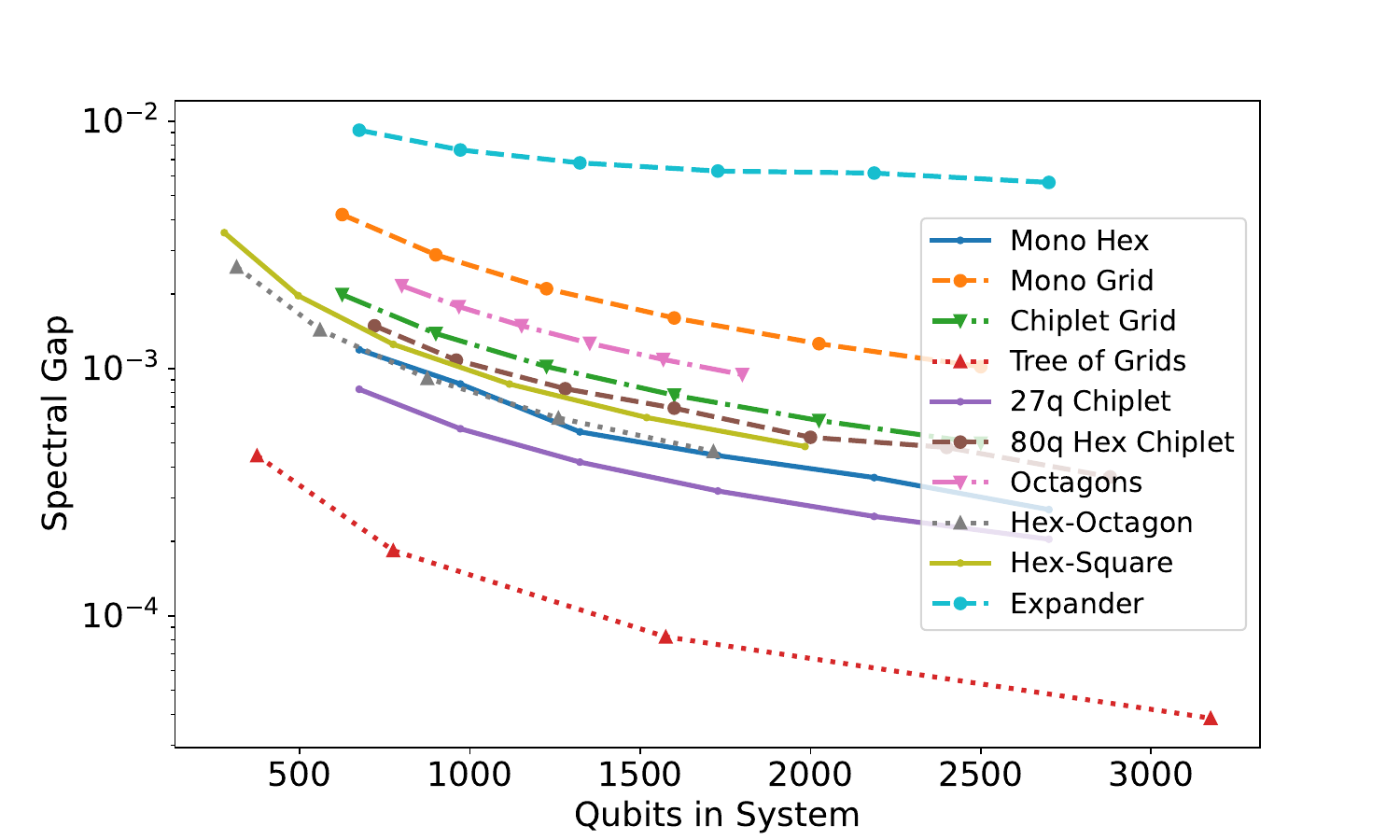}
\caption{Comparison of QC topology spectral gap vs. number of qubits for proposed architectures. The spectral gap as in Definition \ref{def:specgap} quantifies the rate at which random walks on the graph converge toward a uniform distribution on nodes. Graphs with few bottlenecks tend to have higher spectral gaps.}
\label{fig:spect-gap}
\end{figure}

In addition to graph diameter, we consider the spectral gap of a graph to quantify connectivity on a busy network. Here we take the smallest non-zero eigenvalue of the graph's normalized Laplacian matrix as the spectral gap (see \cite[Lecture 7]{williamson2016spectral}, in particular the interpretation given by Cheeger's inequality).
\begin{definition} \label{def:specgap}
For a graph $G$ on $n$ vertices given by its edge set, $G \subset \{(v_1, v_2) \in \{1...n\} \times \{1...n\} \}$, the adjacency matrix $A_G$ is an $n \times n$ matrix given such that $A_{i j} = 1$ when $(i,j) \in G$, and zero elsewhere. The graph's Laplacian matrix is given by $L = \hat{1} - D^{1/2} A D^{1/2}$, where $D$ is the diagonal matrix such that $D_{i i}$ is the degree of node $i \in 1...n$. The spectral gap $\lambda_2$ is the second smallest eigenvalue of $L$.
\end{definition}
As summarized in \cite{hoory2006expander}, a graph's spectral gap relates to the rate at which random walks on that graph approach random node choice. Furthermore, the spectral gap yields bounds on a graph's Cheeger constant, quantifying the ratio between the size of any subgraph and the number of links connecting it to the rest. Hence the spectral gap measures the extent to which a graph has bottlenecks. While graph diameter quantifies the cost of sparse or sequential qubit movements, spectral gap is more relevant when many transfers should occur in parallel. In contrast to graph diameter (for which smaller is better), a large spectral gap is desirable. Computed spectral gaps appear in Figure \ref{fig:spect-gap}.

Some results in Figures \ref{fig:graph-diam} and \ref{fig:spect-gap} are as expected. The monolithic grid architecture is overall more connected than the heavy-hex or chiplet architectures, so it has a lower or comparable graph diameter and higher spectral gap. The tree of grids achieves a good graph diameter but poor spectral gap, because while the hierarchical structure promotes fast routing, it is highly bottlenecked at the root node.

The chiplet grid incurs a larger spectral gap penalty versus its monolithic counterpart than the heavy-hex chiplet architectures. This drop in spectral gap results from the bottlenecks between chiplets. Grid architectures maintain a better ratio of qubits to links, but because of their higher internal connectivity, its not clear that they would more effectively buffer against noise at links or cross-talk over links.

Although topological metrics appear to favor the monolithic grid over other planar layouts, this does not necessarily imply superiority as a quantum computing architecture. As noted in \cite{chamberland2020topological}, heavy-hex and related layouts achieve lower coordination numbers, taking pressure off of other considerations and potentially enabling techniques for higher fidelity.

The difference between distributed and monolithic topologies is often smaller in heavy-hex than in grids. To maintain grid topology would require a link number proportional to the entire chiplet surface area. For the 25-qubit chiplets, this would yield 20 links per chiplet and a qubit-to-link ratio of only 2.5. Since heavy-hex is less internally connected to begin with, one can maintain better qubit-to-link ratios or better isolate the qubits at link endpoints with much smaller disruption to the layout. The chiplet and monolithic architectures based on heavy-hex layouts have similar spectral gaps and graph diameters to each other, which is consistent with expectations. We note that the 80q heavy-hex chiplet sometimes has larger spectral gap than its monolithic counterpart - this small discrepancy probably arises from the extent to which each layout can be made closer to square overall. Applications originally mapped to monolithic grids will likely need to be re-organized around the bottlenecks, while those originally on monolithic, heavy-hex avoid this problem. The lack of an obvious bottleneck in distributed heavy-hex layouts motivates the focus of the next Subsection, in which we consider qubit movements from the perspective of noise.

Finally, we note the combined advantages of expander topologies, which appear to exceed other layouts in spectral gap and graph diameter with a reasonable qubit-to-link ratio, low coordination number, and strong potential isolation of linked qubits. In theory, there exist particularly strong expanders known as Ramanujan graphs \cite{hoory2006expander}. Ramanujan graphs have a constant number of connections per node and constant (non-decaying) spectral gap with total node number. The properties of expanders show how by dropping the commonly assumed constraint of planarity, novel architectures may achieve vast improvements. Expanderized chiplets may show large advantages in quantum error correction using LDPC codes \cite{breuckmann2021quantum}, which often use non-planar connectivity to improve ratios of logical to physical qubits.

\subsubsection{Bandwidth considerations for applications}
The spectral gap may serve as a coarse proxy for an architecture's tendency to avoid bottlenecks in arbitrary computations. In the long-term, a complete analysis of bandwidth likely requires optimizing compilers designed to make best use of connectivity at multiple layers of abstraction, possibly including identifying how problem instances can be mapped to computations that optimize locality of computation. Though chiplets introduce extra heterogeneity, the problem is closely related to that of mapping quantum ciruits to monolithic architectures, in which it is nonetheless useful to minimize overhead of moving data around the device. Though quantum compilation is an active research area, to our knowledge the problem of quantum compilation on large, planar chips remains wide open.

As described in \cite{chamberland2020topological}, error correcting codes exist for heavy-hex and similar layouts designed to minimize the number of connections to each qubit. Such codes are likely to perform comparably on linked architectures such as those shown in Figure \ref{fig:topology-compare} (a) and (b). Though it is beyond the scope of this paper to evaluate error correction thresholds under the heterogeneous noise types induced by chiplets, it is likely that improved on-chip fidelities compensate for the higher likelihood of errors at links.

An alternative strategy is to identify classes of problems that may naturally map to heterogeneous connectivity. In the near-term, chiplet architectures may best suit a class of problems in an intermediate regime between those accessible to classically-connected devices and those requiring densely-connected qunatum architectures. As a point of comparison, some techniques emulate entanglement via classical post-processing \cite{perlin2021quantum, peng2020simulating, eddins2021doubling}. Figures 2 and 3 of \cite{peng2020simulating} suggest how a quantum circuit or physical system might be more amenable than usual to distribution. Good uses of chiplet architectures should require more quantum communication than is tractable to emulate, but possibly less than would general computations.

In entanglement forging \cite{eddins2021doubling} , an observable on a whole system is rewritten as a linear (not necessarily convex) combination of observables on two subsystems via Schmidt decomposition. Entanglement forging simulates the expectation of an observable $\braket{\obs}_{\ket{\psi}}$ on an $n$-qubit system with operator norm error $\epsilon$ via repeated uses of an $n/2$-qubit system. The distributed scheme would use two $n/2$-qubit systems and a number of link uses proportional to the entanglement needed between subsystems. In principle, if each bidirectional link use exchanges up to two qubits of entanglement, the number of link uses is logarithmic in the number of non-zero Schmidt coefficients. For comparability, one should allow a chiplet mapping to approximate the true state within precision $\epsilon$ as well. If we post-select on successful transfers via the two-use code discussed in Section \ref{sec:errdetect}, then we may convert link uses and noise into a failure probability. The chance that each double-transfer fails is $1/\eta$, where $\eta$ is an exponentiated amplitude damping parameter. In this formulation, we can directly compare the number of retrials expected using $\frac{1}{2} \log_2 \Big| \Big \{i : \tilde{\lambda}_i > 0 \Big \} \Big |$ quantum link transfers vs. entanglement forging:
\begin{center} {\small
\begin{tabular}{| c | c |}
\hline
Entanglement Forging & Chiplet Architecture \\ %\hline
 $\frac{1}{\epsilon^2} \Big (  \sum_{k,m} |\lambda_k \lambda_m | \Big )^2$
	& $ \exp \big ( \frac{1}{2} \eta \log_2 |\{i : \tilde{\lambda}_i > 0 \}| \big )$
\\ \hline
\end{tabular} }
\end{center}
If one can tolerate link noise or detect and retry entanglement creation, then the polynomial scaling of retrial numbers in the distributed scheme yields a major advantage in scaling. This advantage is unsurprising, because we expect that for highly entangled systems, a reasonably clean quantum computer should exponentially outperform a classical emulator on tasks involving large, complex entanglement. Applications of quantum pseudorandom states my present such a regime, requiring superlogarithmic but sublinear entanglement across a bipartition \cite{aaronson2024quantum}. Other potential examples may emerge in quantum chemistry \cite{boyn2021quantum-classical}.

\subsection{Modeling qubit movements}
\label{sub:est-fid}
As noted in the previous Subsection, graph-based figures of merit do not capture tradeoffs between higher connectivity and local noise. Indeed, metrics such as graph diameter achieve optima for all-to-all connectivity, as does the ease of designing quantum algorithms and codes. Graph topology is a good way to analyze high-level networks that abstract away problems such as loss and error mitigation, reducing costs to latency, bandwidth, and similar. Graph topology may however be too high an abstraction level to capture important challenges of scaling up quantum computation. As the focus of this paper is on how short-range quantum networks may accelerate the path to scale, quantum advantage, and ultimately fault-tolerance, here we consider noise in both links and local processing to be significant.

Pursuing chiplet architectures may be the most practical route for quantum devices to scale up while keeping noise low. Such may seem counter-intuitive at first glance, because qubits will incur higher noise when traversing links than during usual local gates. In contrast, chiplets may improve local error rates for several reasons:
\begin{itemize}
	\item SC qubits at fixed frequency require individual addressing. Qubits that are too close both physically and spectrally may experience cross-talk in controls.
	\item Unintended coupling between nearby qubits induces cross-talk, especially during operations such as two-qubit gates and readout.
	\item Cosmic rays and other effects induce strongly correlated errors between nearby qubits, potentially evading error correction designed for single-qubit, independent errors \cite{martinis2021saving, mcewen2021resolving}.
	\item Variance in the manufacturing process leads to random fluctuations in chip performance, including occasionally defective, unusable qubits (see \cite[Figure 1]{arute2019quantum} for a dramatic example). Splitting into chiplets could greatly increase the efficiency of defect rejection. For example, if each qubit has a 1\% chance of being defective, the chance of 1000 qubit array emerging with zero defects is roughly $0.4\times 10^{-5}$, likely requiring hundreds of thousands of discarded attempts. As the chance of a 50-qubit array emerging defect-free is about $0.6$, one expects to obtain 20 of these (for 1000 total qubits) in about 34 attempts.
	\item Beyond thousands of qubits per chip, new scaling barriers will likely appear \cite{krinner2019engineering, magnard2020microwave}. As fault-tolerance may need $10^3-10^4$ physical qubits for each logical qubit \cite{fowler2012surface}, achieving this crucial milestone probably requires some combination of redesigning current chips and DRs, major advances in error correction efficiency, and/or distributed computing.
\end{itemize}
A recent meta-analysis \cite{sevilla2020forecasting} begins to quantify the tradeoff between qubit number and noise rates. These authors estimate the correlation between qubit number and error as within the interval $(-0.11, 0.48)$ with a confidence of 0.84 that the correlation is positive. In other words, as monolithic devices increase in size, it is likely that the average fidelity of two-qubit operations will decrease. In contrast, small chips joined by short wires could avoid or reduce many of these penalties to scaling.

In this Subsection, we combine ideas from the physical channels models described in Subsection \ref{sub:chanmod} with the network modeling of the previous Subsection \ref{sub:graph-diameter} and the aforementioned chip size vs. noise tradeoffs. We study the noise incurred by a single qubit traversing the graph diameter, calculating the impact of each local SWAP gate or microwave link. We focus on the 27-qubit Falcon processor as chiplet model as illustrated in Figure \ref{fig:topology-compare} for several reasons.
\begin{enumerate}
\item As described in Subsubsection \ref{sec:swchainchan}, we have access to and can analyze IBM Falcon processors in detail.
\item This layout attaches links to qubits that each have only a single local connection. Because linking qubits are reasonably well-isolated and at the physical edge of chiplets, we make the simplifying assumption that chiplet internal SWAP gate noise does not increase with the number of chiplets in a system, only with the number of locally-connected chips.
\item Though the relatively small ratio of qubits to links is a potential caveat to the previous point, we believe this is accounted for in the analysis. Since we count extra noise incurred at each link, high link density results in pessimistic estimates. Future engineering considerations will probably favor larger chiplets, such as the 127-qubit IBM Eagle processor. We expect the metrics herein to be more favorable for larger chiplets, since they would require fewer link crossings. Similarly, expander-based graphs should achieve vastly improved performance. Hence we take the linked Falcon architecture as pessimistic on the metrics we quantify and relatively safe on those we cannot.
\item As concluded in the previous Subsection \ref{sub:graph-diameter}, similarity between the chiplet and monolithic architectures motivates analyzing this layout entirely in terms of noise incurred during qubit movements. In this Subsection, we do not directly address bandwidth or latency, which should be only slightly worse than in monolithic heavy-hex.
\end{enumerate}

\subsubsection{A simple model: decaying fidelity}
\label{sub:est-fid-near}
We aim to model the process fidelity of state transfer cross-system, or the graph diameter path, using values based on today's real machines, reported trends, and thresholds for error correction. A logical SWAP operation between two qubits allows qubit information to move cross-chip, and as seen in Figure~\ref{fig:swap-CX}, a SWAP can be decomposed into three CX operations. The IBM quantum gate library includes the two-qubit CX, so the Figure~\ref{fig:swap-CX} circuit is commonly used for communication between non-local qubits. Thus, average CX fidelity is an integral part of state transfer process fidelity analysis.

A recent, public competition sought pulse-level controls to minimize SWAP gate noise, potentially replacing the standard 3-CX implementation \cite{lanes2021announcing}. Unfortunately, crowdsourced solutions did not exceed the 50\% improvement standard set by the competition organizers. Therefore, the 3-CX implementation still represents a good if rough model of how local SWAPs work on real devices.

Here we propose a noise model based on standard, simplifying assumptions, in particular that noise is typically depolarizing. As noted in Subsection \ref{sub:chanmod}, the depolarizing noise assumption is not physically realistic for current microwave links, which primarily induce amplitude-damping noise. It is more reasonable for local SWAP gates and compatible with the randomized benchmarking techniques used by IBM to obtain average CX gate errors as reported ~\cite{randBench,magesan2011scalable,magesan2012characterizing}. It is not physical realism but a simplifying abstraction that we seek here. In the next Subsubsection, we will address the physical picture in much greater detail.

The basic tradeoff assumes that as the number of locally connected qubits increases, the fidelity of local operations falls. To model the direct relationship between monolithic CX gate fidelity, $F_{CX,mono}$, and qubit count, $N$, we calculate monolithic CX fidelity as
\begin{equation} \label{eq:mono-calc}
\begin{split}
    F_{CX,mono} & = 1-((N-n_{chip}) \times \Delta_{infid.} \\  + & (1-F_{CX,chip}))
\end{split}
\end{equation}
for intra-device operations on the monolithic QC where $\Delta_{infid.}$ is average gate infidelity increase per additional monolithic-system qubit above the number of qubits in one chiplet, $n_{chip}$. Each 27-qubit chip modeled after a Falcon processor is initially assigned the two-qubit gate fidelity of $F_{CX, chip}=F_{CX,Mumbai}$. $F_{CX,mono}$, as defined in Equation~\ref{eq:mono-calc}, is calculated using $F_{CX,chip}$ and the total number of qubits in the system, $N$.

$F_{CX}$ and $F_{link}$ values described above are used to approximate total process fidelity of cross-system state transfer on various sizes of the monolithic and chiplet architectures. Process fidelity is approximated with average infidelity as outlined in ~\cite{sanders2015bounding}. Although there is no provably direct connection between average gate fidelity and error rate~\cite{sanders2015bounding}, approximation of the combined noise process for each operation is feasible. Using methods found in~\cite{sanders2015bounding}, we estimate the total noise, $r$, at each step in the SWAP chain before values are combined to approximately model the success of a cross-system SWAP chain. If we assume that all noise is depolarizing, then $r$ is the noise parameter under the depolarizing channel in dimension $d$ given by
\[\rho \mapsto (1-r) \rho + r \frac{\hat{1}}{d} .\]
As depolarizing noise eventually reduces the state to a complete mixture rather than one orthogonal to its original value, the relationship between $r$ and fidelity depends on dimension. We will also work with the non-depolarizing parameter $R := 1 -r$. Qubit state transfer, a single qubit operation, over a microwave link is modeled with the error parameter
\begin{equation}
    r_{link} = 2 \times (1-F_{link}), R_{link} = 1-r_{link} .
\end{equation}
An on-chip CX operation, as a two-qubit operation, is modeled with the error parameter
\begin{equation}
    r_{CX} = \frac{4}{3} \times (1-F_{CX}), R_{CX} = 1-r_{CX}.
\end{equation}
An on-chip SWAP is comprised of three CX operations, as pictured in Figure~\ref{fig:swap-CX}. Thus,
\begin{equation}
    R_{SWAP} = (R_{CX})^3.
\end{equation}
The formula does not change after tracing out one of the qubits. In our simplified model, a noisy qubit input to a SWAP gate retains any initial noise as it is transferred to its new location, so its final (single-qubit) noise multiplies its initial non-depolarizing ratio by the two-qubit non-depolarizing ratio of the SWAP gate. The primary convenience of this model is that non-depolarizing ratios multiply all the way down a chain. $R_{total}$ is determined by taking the product of $R_{link}$ or $R_{SWAP}$ values along the system's longest path. Finally, total process fidelity is approximated using the correction
\begin{equation} \label{eq:simplefid}
    F_{process} = 1 - \frac{1-R_{total}}{2}.
\end{equation}
Though this model begins to incorporate some physical effects of qubit movements, it is still highly abstracted, ignoring qualitative distinctions between types of noise. In the next Subsubsection, we we validate this model by comparing to one of much greater physical realism.

\subsubsection{Modeling qubit movements via physically inferred channels} \label{sec:swchainchan}
We seek to address two primary concerns lingering after the previous analyses. First, while analyzing the fidelity of moved qubits gives a broad picture of where distributed quantum computing could show advantages over monolithic, efforts to evaluate specific algorithms or applications might need to consider specifically what kind of errors occur in each setting. In Section \ref{sec:errdetect}, we use the more detailed models to suggest error detection for links.

Second, we hope to understand whether abstractions are useful rather than hiding important aspects. As in Section \ref{sub:graph-diameter}, models based purely on network topology favor as much connectivity as possible. We know however that there are constraints and tradeoff. Section \ref{sub:est-fid-near} begins to address these tradeoffs, incorporating them into a model that shows where distributed and monolithic architectures may compete. We still however must acknowledge that the fidelity calculations ignore qualitative differences in the kind of noise from SWAP gates and links. In this Section, we examine the more physically detailed models underlying and abstracted by the fidelity calculations. The results herein validate those abstractions while illuminating their limitations.

In principle, the one-qubit link and local SWAP channels given by Choi matrices in Equations \eqref{eq:oneswapchoi} and \eqref{eq:linkchoi} should compose to the channel undergone by a qubit moving through any sequence of local SWAP operations and links. In practice, several effects limit the accuracy of such a procedure: heterogeneity in the device, variation of parameters over time, inability to fully extract readout errors from local SWAP and link tomographies, etc. In practice, we find that a different model yields a better combination of accuracy and simplicity.

\begin{figure}[t!]
\includegraphics[width=0.4\textwidth]{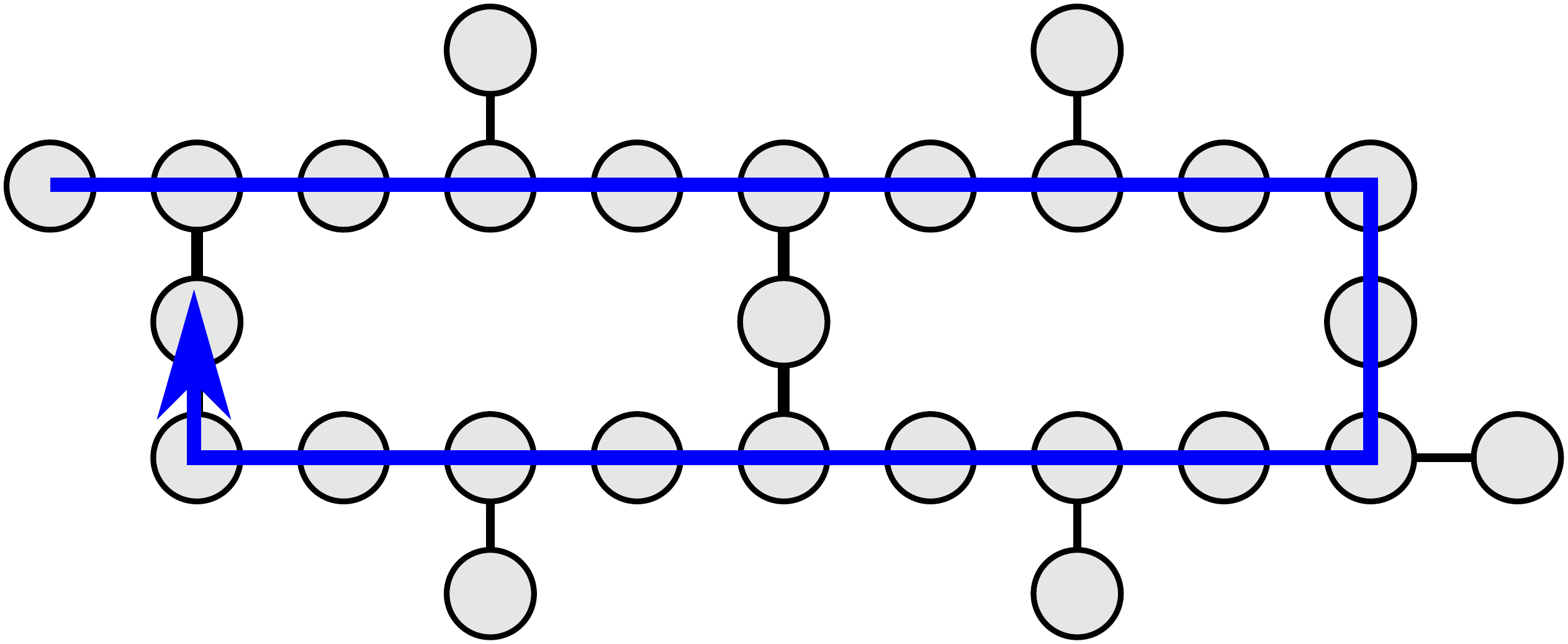}
\caption{Path of SWAP gate chains through IBM Falcon processors.}
\label{fig:falc-path}
\end{figure}

\begin{figure*}[!tb] \scriptsize \centering
	\begin{subfigure}[b]{0.33\textwidth}
		\includegraphics[width=0.99\textwidth]{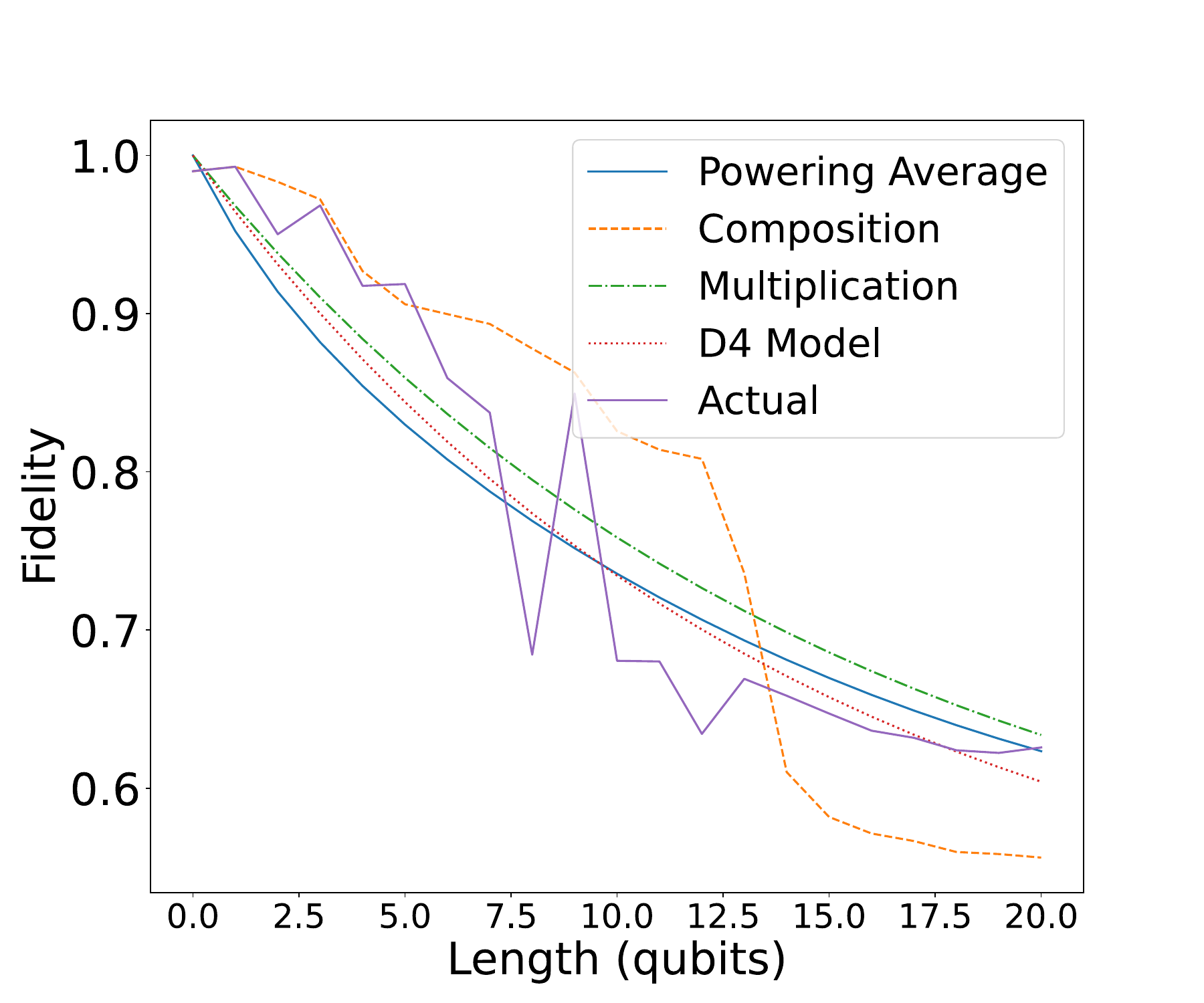}
		\caption{}
	\end{subfigure}
	\begin{subfigure}[b]{0.33\textwidth}
		\includegraphics[width=0.99\textwidth]{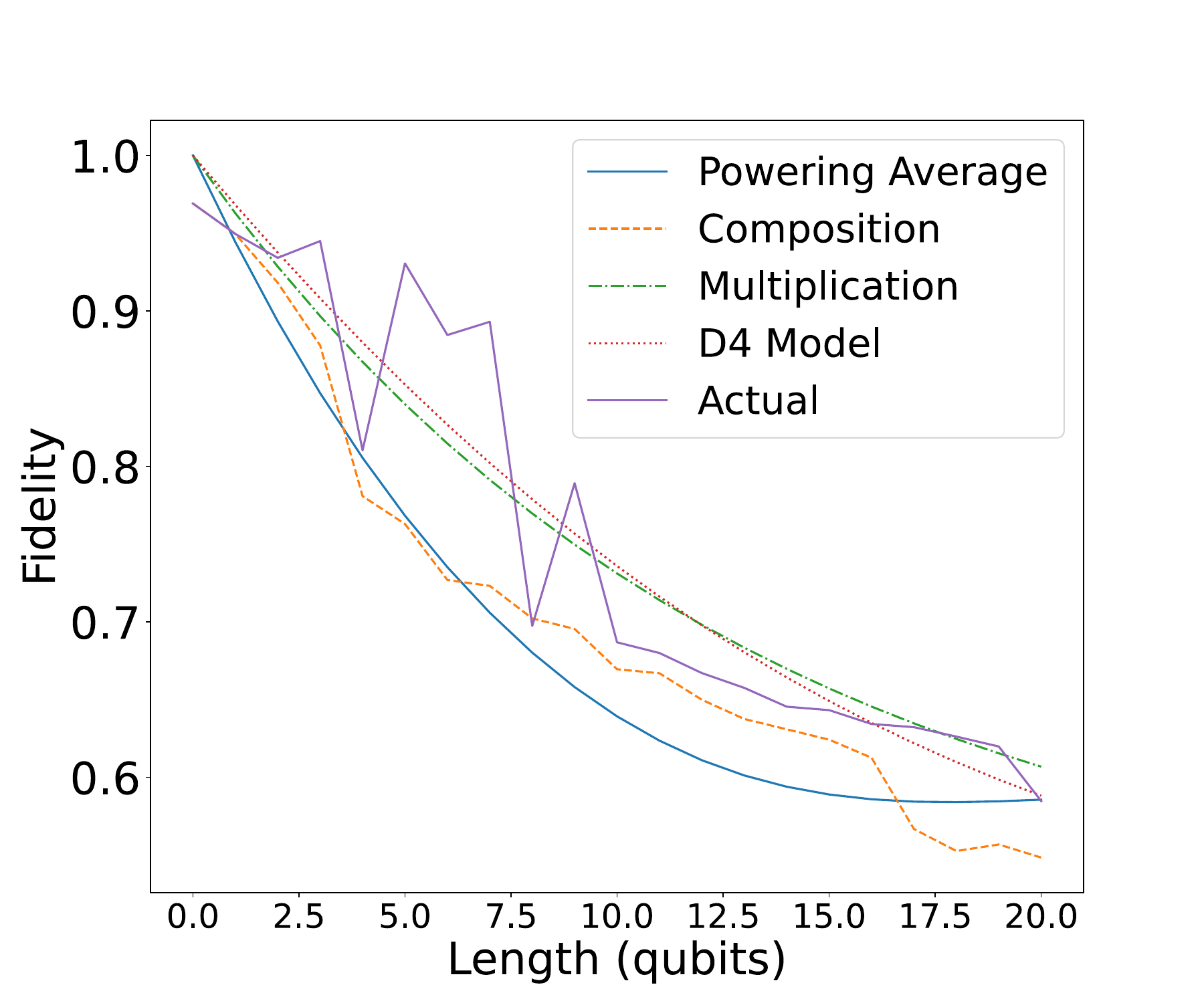}
		\caption{}
	\end{subfigure}
	\begin{subfigure}[b]{0.33\textwidth}
		\includegraphics[width=0.99\textwidth]{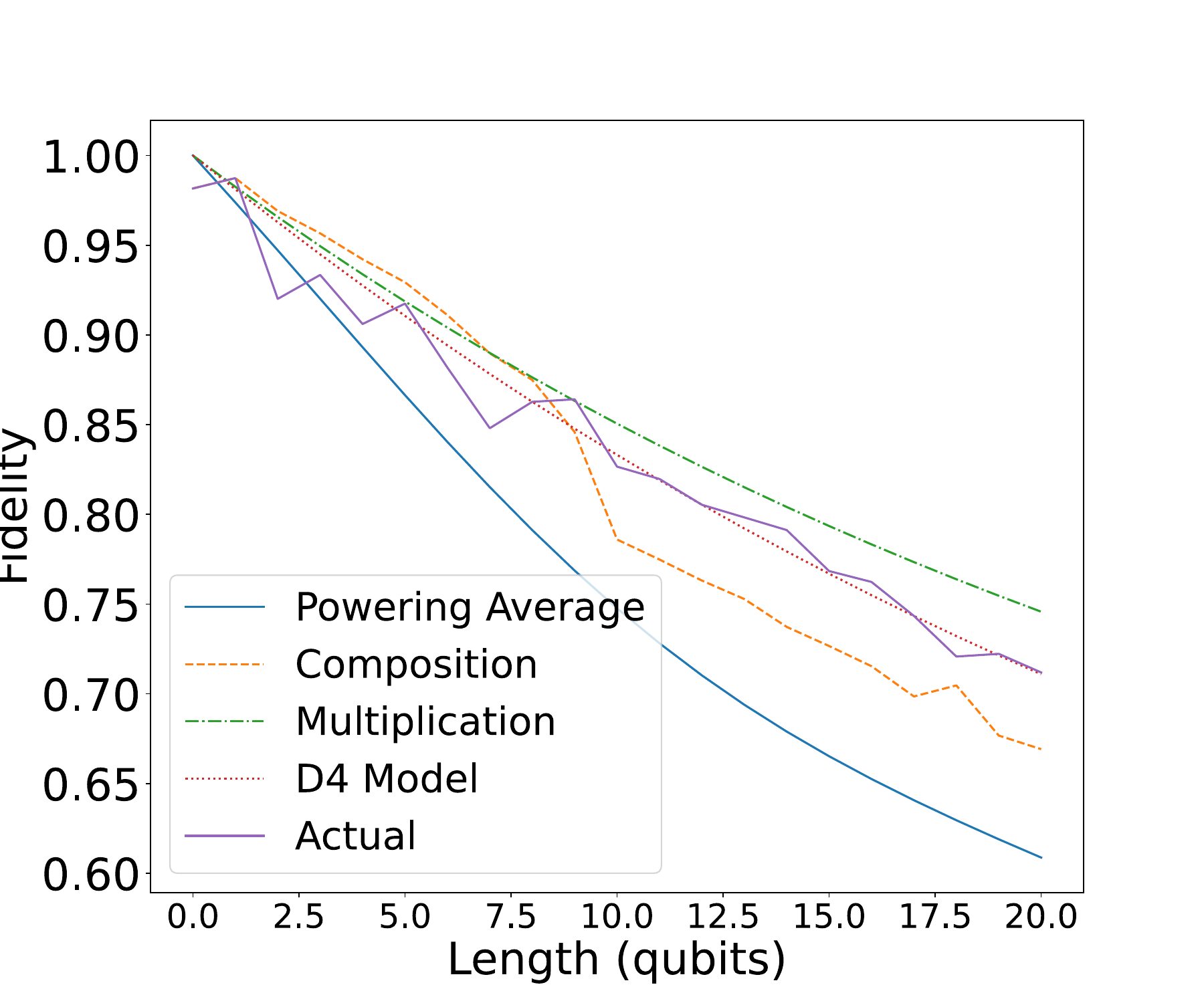}
		\caption{}
	\end{subfigure}
	\caption{Computed output fidelity with the identity process by the 5 chosen methods from (a) \textit{ibmq\_montreal} (b) \textit{ibmq\_sydney} (c) \textit{ibmq\_mumbai}. \label{fig:swapchainfid}}
\end{figure*}

To infer the channel that a moving qubit undergoes from local SWAP operations, we apply a one-qubit channel tomography with preparation on the starting qubit and measurements on the final location. We consider SWAP chains starting at length zero (qubit 0 is prepared, then measured), and going to length 20 (a qubit is prepared, then a chain of SWAP gates moves its state via a long route, which is measured). The particular path is shown in Figure \ref{fig:falc-path}. We obtain expected output fidelities by four methods, as shown in Figures \ref{fig:swapchainfid}. Channel tomographies were performed using IBM Qiskit's built-in process tomography routines~\cite{qiskit-tomo}. We then applied a relatively simple form of readout error mitigation, first applying the inverse matrix calculated (see \cite{noauthor_measurement_nodate-1}). To ensure that the matrix would be positive and normalized, we first added complete mixture to counteract any negative terms, then scaled the matrix by its inverse trace.

We also initially considered SWAP chains on the \textit{ibmq\_brooklyn}, a 65-qubit processor allowing longer chains. Unfortunately, the longer chains encountered constraints on the number of circuits executable within a single job via IBM's cloud interface. Because they were split across jobs, these produced anomalous results due to temporal fluctuations in device parameters.
%In particular, we study SWAP chains on the 65-qubit \textit{ibmq\_brooklyn} and 27-qubit \textit{ibmq\_montreal}. Properties of these devices as well as their layouts with physical qubits labeled can be found on the IBM quantum systems page~\cite{IBMQS}. For the \textit{ibmq\_brooklyn}, we take a contiguous path of qubits given by indices [0, 1, 2, 3, 4, 5, 6, 7, 8, 12, 21, 20, 19, 18, 17, 16, 15, 24, 29, 30, 31, 32, 33, 34, 35, 40, 49, 48, 47, 46, 45, 44, 43, 52, 56, 57, 58, 59, 60, 61, 62, 63, 64].

In truth, moving a qubit through the device touches every qubit crossed. We may however assume roughly that each qubit on the path undergoes the noise associated with a single SWAP as modeled by Equation \eqref{eq:duplexswapchoi}. Though this simplification requires some assumptions, it avoids having to do expensive multi-qubit tomographies. A 20-qubit tomography would not be tractable on today's devices as the number of required circuits scales exponentially with number of qubits.

We consider five ways of calculating the fidelity at the end of a SWAP chain, including four that calculate the eventual Choi matrix, here listed in order of increasing abstraction.
\begin{enumerate}
	\item \textit{Actual}: SWAP chain channel tomography, treating the SWAP-chain as a one-qubit channel as described in Subsection \ref{sub:chanmod}.
	\item \textit{Composition}: individual channel tomographies are performed on pairs of neighboring qubits. We perform multiple trials to get most pairs, leaving one idle qubit between simultaneously evaluated pairs to reduce readout cross-talk. One-qubit SWAP channels are composed to estimate the channel of the full chain.
	\item \textit{Powering Average}: the average of the one-qubit SWAP channels is taken to the power corresponding to the chain length. This method is less detailed than the above channel composition method, because the averaged channel erases knowledge about heterogeneity of different qubits on the same device.
	\item \textit{\dfour Model}: described below.
	\item \textit{Multiplication}: calculated as in Section \ref{sub:est-fid-near}.
\end{enumerate}
Noise in real systems often includes mixing, relaxation, and unintended rotation. We capture most of the noise we observed in what we call the \textit{Depolarizing-Damping-Dephasing-Drift} (\dfour) model. Depolarizing, amplitude-damping, and dephasing noise are respectively described for a qubit density $\rho$ by the quantum channels:
\begin{equation} \label{eq:ddd}
\begin{split}
& \Phi_{\text{depo}(a)}(\rho) 
		= (1-a) \rho + a \begin{pmatrix} 1/2 & 0 \\ 0 & 1/2 \end{pmatrix} \\
& \Phi_{\text{damp}(a)} \begin{pmatrix} \rho_{1 1} & \rho_{1 2} \\ \rho_{2 1} & \rho_{2 2} \end{pmatrix}
	= \begin{pmatrix} \rho_{1 1} + a \rho_{2 2} & \sqrt{1-a} \rho_{1 2}
		\\ \sqrt{1 - a} \rho_{2 1} & (1 - a) \rho_{2 2} \end{pmatrix} \\
& \Phi_{\text{deph}(a)} \begin{pmatrix} \rho_{1 1} & \rho_{1 2} \\ \rho_{2 1} & \rho_{2 2} \end{pmatrix}
		= \begin{pmatrix} \rho_{1 1} & (1-a) \rho_{1 2} \\ (1-a) \rho_{2 1} & \rho_{2 2} \end{pmatrix} .
\end{split}
\end{equation}
Here $a$ is a parameter between 0 and 1 controlling the noise strength. In real chains of SWAPs on the IBM QCs, we also noticed a rotation contribution of the form
\begin{equation} \label{eq:rot}
\Phi_{\text{rota}(\theta)}  \begin{pmatrix} \rho_{1 1} & \rho_{1 2} \\ \rho_{2 1} & \rho_{2 2} \end{pmatrix}
	= \begin{pmatrix} \rho_{1 1} & e^{i \theta} \rho_{1 2} \\ e^{- i \theta} \rho_{2 1} & \rho_{2 2} \end{pmatrix} .
\end{equation}
Since amplitude damping and depolarizing noise do not commute, there is some ambiguity in how one models a quantum process that imposes both. We take the combined channel model corresponding to the continuous limit of both noise types occurring simultaneously. This approach yields a quantum Markov semigroup, a family of channels parameterized continuously by time. %The obtained form has a unique and usually full rank fixed point state as long as depolarizing strength is non-zero, a useful property in theoretical analysis and in estimating rates of decay to equilibrium \cite{bardet2017estimating, carlen2017gradient}. While the \dfour form contains some unitary rotation, because this rotation commutes with the restriction to the fixed point algebra, one may still apply recent results showing exponential decay of relative entropy to the fixed point \cite{junge2019stability, gao2021complete, gao2021geometric}.

We let $\Phi_{D(\epsilon, \eta, \delta, \theta)}$ denote the qubit channel given by the \dfour model, where $\epsilon$ is the strength of depolarizing noise, $\eta$ is the strength of amplitude-damping, $\delta$ is the strength of dephasing, and $\theta$ is the rotation angle. Let $\rho_{44}^{(eq)} = \epsilon / 4(\epsilon + \eta)$. A quantum channel in finite dimension $d$ is fully characterized by its Choi matrix, the resulting density from the input of a $d \times d$ Bell pair through one side of the channel. The Choi matrix of $\Phi_{D(\epsilon, \eta, \delta, \theta)}$ is then given by
\begin{widetext}
\begin{equation}
\left(\begin{matrix}
\frac{1}{4}(1 + e^{-\epsilon t}) & 0 & 0 & \frac{1}{2} e^{-\epsilon - \eta/2 - \delta + i \theta} \\
0 & \frac{1}{4} (1 - e^{-\epsilon t}) & 0 & 0 \\
0 & 0 & \big (\frac{1}{2} - \rho_{44}^{(eq)} \big) \big (1 - e^{-(\epsilon + \eta) t} \big) & 0 \\
\frac{1}{2} e^{-\epsilon - \eta/2 - \delta - i \theta}  & 0 & 0 & \rho_{44}^{(eq)} + \big (\frac{1}{2} - \rho_{44}^{(eq)} \big) e^{-(\epsilon + \eta) t} \\
\end{matrix} \right).
\end{equation}

\begin{table}
\centering
\begin{tabular}{| c || c | c | c | c || c | c | c || c |} \hline
Device & $\epsilon$ & $\eta$ & $\delta$ & $\theta$ & Pow & CC & \dfour & Fid \\ \hline
\textit{montreal} & 0.048 & 0.013 & 0.026 & -0.024 & 0.82 & 0.94 & 0.96 & 0.93 \\ \hline
\textit{sydney} & 0.052 & 0.004 & 0.014 & -0.056 & 0.83 & 0.94 & 0.95 & 0.93 \\ \hline
\textit{mumbai} & 0.030 & 0.008 & 0.005 & -0.025 & 0.82 & 0.96 & 0.97 & 0.94 \\ \hline
\end{tabular}
\caption{\dfour parameters and reconstruction fidelities for swap chain tomographies on 3 IBM devices. As column headings, ``Pow," ``CC," and ``\dfour" refer respectively to the process fidelities of reconstructed Choi matrices using the powering average, composition, and \dfour methods as described above. The last column labeled ``Fid" contains the average percent correctness simplified fidelity calculation as in Equation \eqref{eq:simplefid}, where percent correctness is given by one minus the difference between calculated and actual fidelity over the actual fidelity. All data in this table were taken on November 29, 2021 remotely via IBM Qiskit. \label{tab:swapchains}}
\end{table}
\end{widetext}
In common notation, $\eta = 1/T_1$, the relaxation time, while $\delta = 1/T_2$ is the dephasing time. The equilibrium population of the $\ket{11}$ state, given by $\rho_{44}^{(eq)} = \epsilon / 4(\epsilon + \eta)$, arises from the tradeoff between depolarizing and damping noise in the long time limit. The \dfour channels corresponding to different depolarizing vs. damping noise levels will not necessarily commute with each other. Since we will usually consider discrete SWAP gates and link uses, we treat the ``time" parameter $t$ as discrete, counting the number of identical SWAP gates or link uses. Here we may think of the parameters $\epsilon, \eta, \delta, \theta$ has having natural units of the inverse time needed for one basic qubit movement.

Listed in Table \ref{tab:swapchains} are the fidelities of reconstructed Choi matrices against actual from channel tomographies. We observe:
\begin{itemize}
	\item As expected, the primary contribution to irreversible SWAP gate noise is depolarizing.
	\item There is a substantial contribution from coherent phase drift. With calibration, it might be possible to correct this drift by applying phase gates or pulses after SWAP and even CX gates.
	\item The \dfour model obtains average channel reconstruction fidelities above 95\%. Unlike channel composition, the \dfour model enables simple extrapolation to longer chains. The powering average method also naturally extrapolates but yields noticeably worse reconstructions.
	\item The simplified fidelity calculation as in Equation \eqref{eq:simplefid} is more than 90\% accurate on average despite ignoring the distinction between qualitatively different errors.
\end{itemize}
These SWAP chain results motivate the simplified fidelity from Equation \eqref{eq:simplefid} and the \dfour model as well balanced between capturing enough physics to be realistic and abstracting enough detail to simplify calculations.
In Figure \ref{fig:swapchainfid}, we show the calculated and actual fidelities of swap chains with the identity channel on 3 IBM Falcon processors.

We may include links in a \dfour or channel composition model of a SWAP chain. For a hypothetical device, we should have a simple way to extrapolate beyond observed qubits, which favors a linearized \dfour or powered average approach. Based on accuracy and simplicity, the \dfour approach actually appears stronger. Since the link and local SWAP channels in the \dfour model probably do not commute with each other, we cannot simply add up parameters. The matrix multiplications are however easy numerically. A simple model for a long link-SWAP chain across many chiplets is given by
\begin{equation} \label{eq:dddrcompose}
\Phi^k = \big (\Phi_{\text{D}(\epsilon, \eta, \delta, \theta)}^m \Phi_{\text{D}(0, \tilde{\eta}, 0, 0)} \big )^k ,
\end{equation}
where $m$ is the number of local SWAP gates before traversing each link, $k$ is the total number of chiplets traversed, and $\tilde{\eta}$ is the damping parameter of the link. This model is actually very reasonable on highly regular architectures. For example, a one-dimensional line of connected grids would have $m$ equal to the width of each chiplet, and $k$ equal to the total chiplet number. For a corner-to-corner path on a two-dimensional grid, we would roughly set $m$ equal to width plus height of each rectangular chiplet, and $k$ equal to the number of chiplets across two sides.

Comparing the \dfour model to the simplified model of Equation \eqref{eq:simplefid}, we find reasonable agreement. We consider the difference in estimated fidelity across values of $m$ and $k$ each ranging from 1 to 20, \dfour parameters from each row of Table \ref{tab:swapchains} and a link with $\eta = 0.12$. The largest difference between fidelities calculated via \dfour and Equation \eqref{eq:dddrcompose} remains below 0.045, which is below 10\% of the estimated value from either. Ultimately, the results of this Subsection justify the use of Equation \eqref{eq:simplefid} in broadly estimating fidelities. Nonetheless, the more detailed \dfour model will likely be useful in fine-grained analysis and designing applications. As discussed in Subsection \ref{sec:errdetect}, knowing the qualitative form of noise enables simplified and improved error handling.

\subsubsection{Chiplet vs. monolithic performance on qubit movements}
\label{sub:est-fid-res}
In this Section, we compare the fidelity of a qubit that has been moved around a device on monolithic against a chiplet architecture.
On September 7, 2021  the 27-qubit \textit{ibmq\_mumbai} reported an average CX fidelity of approximately $F_{CX, Mumbai}=1-0.009$ while the 65-qubit \textit{ibmq\_brooklyn} reported  $F_{CX, Brooklyn}=1-0.018$~\cite{IBMQE}. This comparison likely represented a much better day for the \textit{ibmq\_mumbai} than for the \textit{ibmq\_brooklyn} and is not a statistically rigorous expectation of typical values. It nonetheless gives a pessimistic estimate of $\Delta_{infid.}$, which we take as an upper bound on how infidelity of monolithic processors may scale.
\begin{equation}
\begin{split}
\frac{F_{CX,Mumbai}-F_{CX,Brooklyn}}{N_{Brooklyn} -N_{Mumbai}} & = \frac{0.991-0.982}{65-27}
 \\ & \approx 0.0002.
\end{split}
\label{eq:infid-rate-calc}
\end{equation}
As a lower bound, we may assume $\Delta_{infid.} = 0$, corresponding to no excess noise for larger chips. Since our analysis specifically considers Falcon processors, a class of processors seen online from 2020-2024~\cite{smith2023fast,IBM-retired-processors}, we do not attempt to incorporate the broader numerical estimates from \cite{sevilla2020forecasting}. Our bounds on $\Delta_{infid.}$ are coarse, representing high uncertainty. Rather than attempt to make detailed predictions about forthcoming technologies, we aim to capture a wide range of possibilities.

We consider three cases for microwave state transfer fidelity between chips in our proposed systems based on the state-of-art. For the purpose of this analysis, we will treat $F_{link}$ as the average gate fidelity inter-chip,  single-qubit state transfer. First, the lower bound in is assigned $F_{link}=0.86$ as reported in~\cite{magnard2020microwave}. The recent demonstration in~\cite{zhong2021deterministic} serves as our second point of comparison for link performance with a value of $F_{link}=0.91$. It is suggested that state transfer fidelity could reach as high as 0.96 as processes are refined~\cite{magnard2020microwave}, so $F_{link}=0.96$ serves as the upper bound for link performance.

\begin{figure}[h!]
\includegraphics[width=0.49\textwidth]{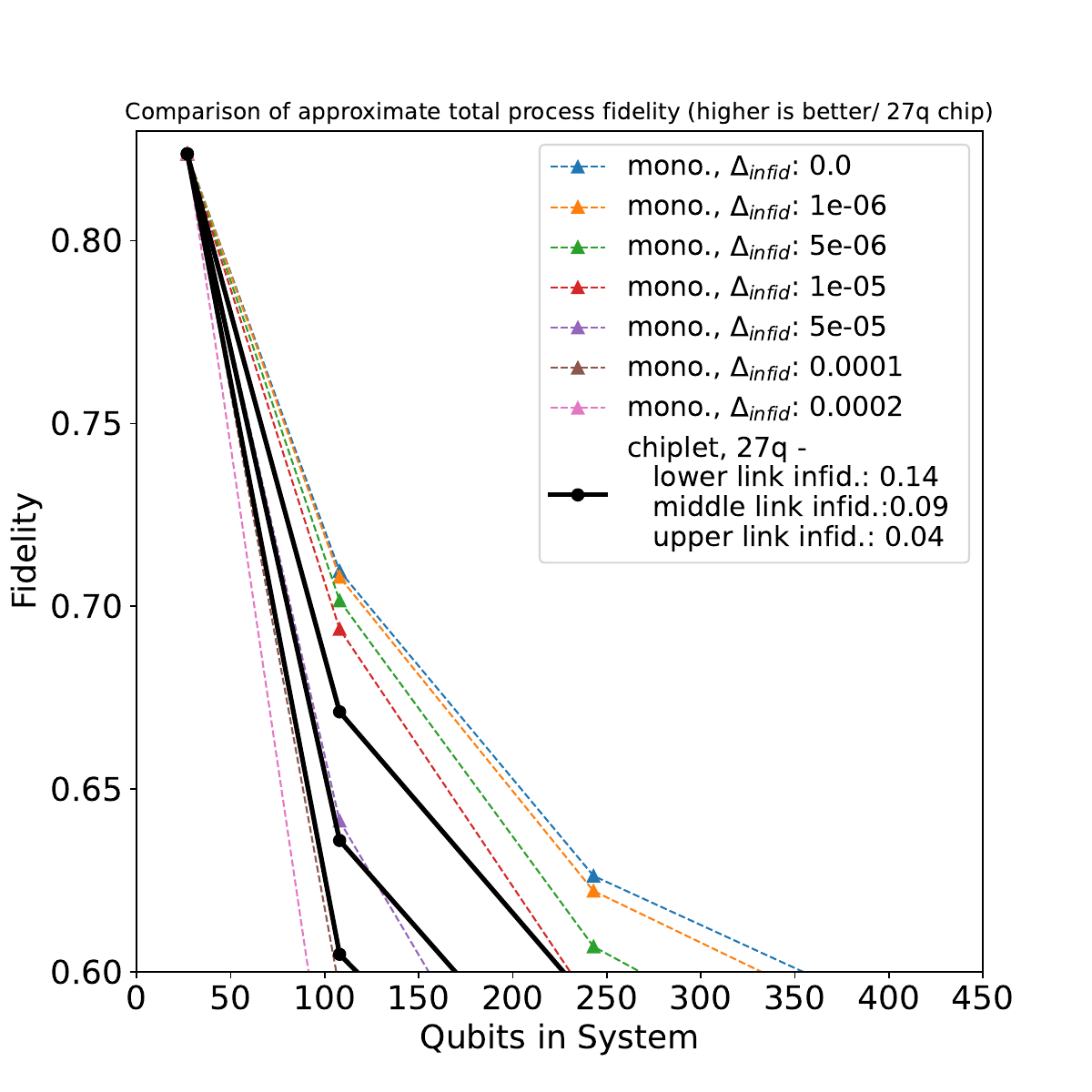}
\caption{Plots approximating total process fidelity of qubit state traversing a system's graph diameter. Proposed 27-q chiplet architectures with $F_{CX,chip}=0.991$, $F_{CX,Mumbai}$  are compared to monolithic architectures. The black curves define a space where modeling predicts chiplet architecture performance depending on the quality of the link (lower boundary: $F_{link}=0.86 $/ middle curve: $F_{link}=0.91$/ upper boundary: $F_{link}=0.96$). Various monolithic curves coarsely approximate different trends in monolithic, two-qubit gate infidelity as devices scale past the size of the chiplet (27 qubits). }
\label{fig:graph-fid}
\end{figure}

Results are featured in Figure~\ref{fig:graph-fid}. Figure~\ref{fig:graph-fid} includes a comparison of total process fidelity of qubit state transfer on the 27-qubit chiplet architecture  vs. monolithic. When examining process fidelity of similarly-sized systems containing more than one chiplet, the both architectures perform similarly.

The black curves of Figure~\ref{fig:graph-fid} enclose a space where our model predicts chiplet architecture performance depending on the quality of the link. The black curve forming the lower boundary of this space represents a chiplet architecture as it increases in size with $F_{CX,chip}=F_{CX,Mumbai}=1-0.009$ and $F_{link}=0.86$. These values are constant even as system sizes increase under the assumption that modular elements of a chiplet architecture can be consistently manufactured to meet design specifications that require minimum performance thresholds. The middle black curve represents a chiplet architecture as it increases in size with $F_{CX,chip}=F_{CX,Mumbai}$ and $F_{link}=0.91$. The upper boundary of the chiplet process fidelity space represents the optimistic outlook where $F_{CX,chip}=F_{CX,Mumbai}$ and the link transfer fidelity is equal to $F_{link}=0.96$. We anticipate future improvements in link technology will allow network performance to fall within the bounds of the chiplet region of Figures~\ref{fig:graph-fid}(a-b), moving closer to the top over time. As a note, our analysis here holds $F_{CX,chip}$ constant as 1) a reference point for $F_{CX,mono}$ and 2) a minimum average gate fidelity target for mass produced QC chiplets. Future developments will likely improve $F_{CX,chip}$ as well.

Included in Figures~\ref{fig:graph-fid}(a-b) are various monolithic curves that coarsely approximate SWAP chain process fidelity according to different trends in monolithic, two-qubit gate infidelity as devices scale past the size of the chiplet (27 qubits). The top, light-blue curve represents the optimistic outlook where two-qubit gate infidelity of monolithic devices consistently match chiplet two-qubit infidelity (i.e. $1-F_{CX,mono}=1-F_{CX,chip}$) at all $N$. This is an unlikely scenario. The remaining curves for monolithic SWAP chain fidelity see the anticipated, negative correlation between number of qubits and two-qubit average gate fidelity. Each curve represents a different average gate infidelity increase, $\Delta_{infid.}$, per additional monolithic-system qubit. The lowest pink curve, is a pessimistic outlook where the infidelity trend scaling from the 27-qubit to 65-qubit machine is extrapolated.

One might ask if potential chiplet advantages extend to better local processors, as future devices are expected to exceed the performance of the \textit{ibmq\_mumbai}. For a more optimistic estimate of local error, we will consider a relevant error correction threshold. A hybrid surface and Bacon-Shor error correcting code with an error threshold of approximately 0.0045 is targeted for use on the heavy-hex lattice~\cite{chamberland2020topological}. As a note, surface code has an error tolerance of approximately 0.01 while traditional Bacon-Shor codes require errors lower than $2\times10^{-5}$~\cite{fowler2012surface}.

\begin{figure}[h!]
\includegraphics[width=0.49\textwidth]{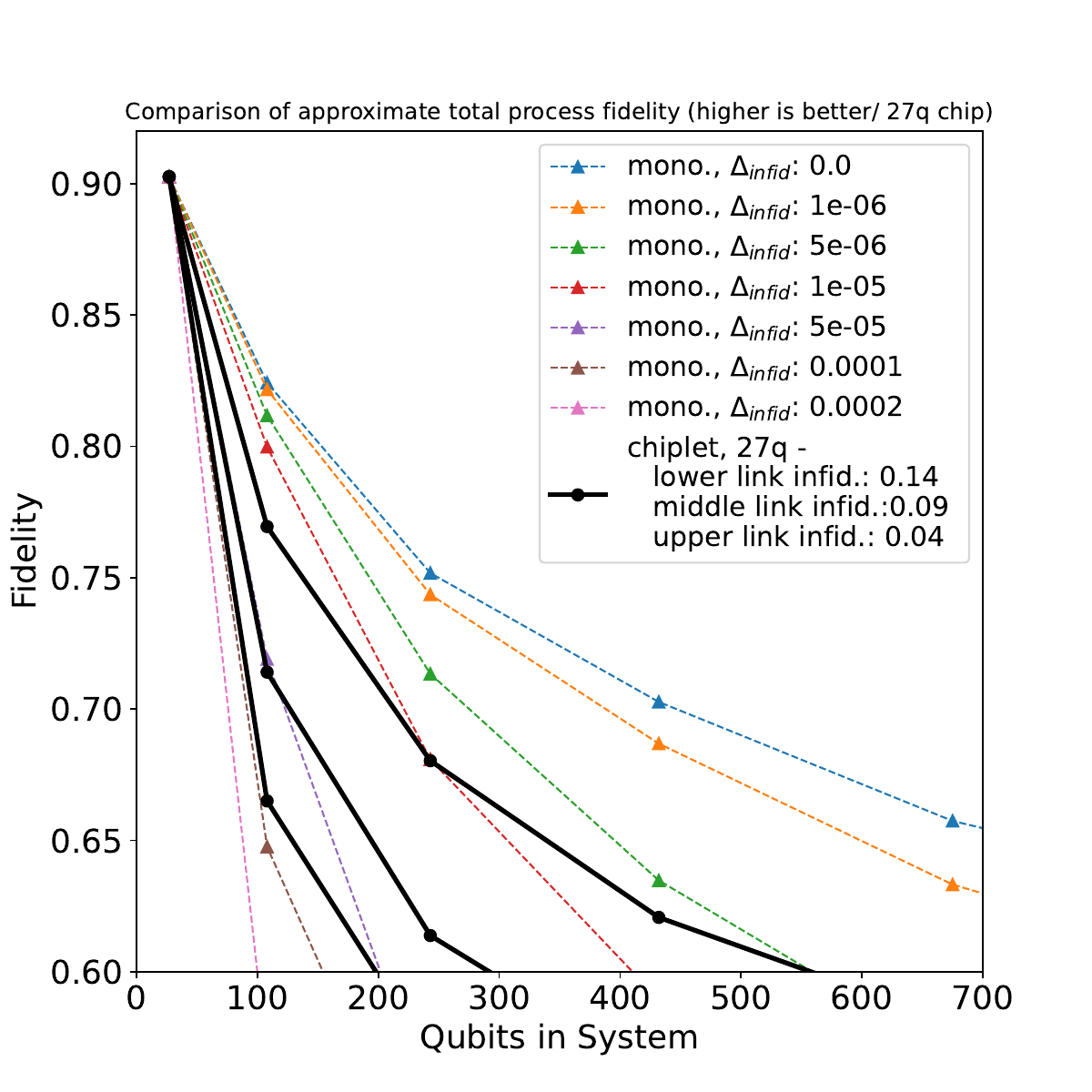}
\caption{Plots approximating total process fidelity of qubit state traversing a system's graph diameter on proposed 27-q chiplet and architecture with $F_{CX,chip}=0.9955$, the hybrid surface/Bacon-Shor threshold, are compared to monolithic architectures. The black curves define a space where modeling predicts chiplet architecture performance depending on the quality of the link (lower boundary: $F_{link}=0.86 $/ middle curve: $F_{link}=0.91$/ upper boundary: $F_{link}=0.96$). Various monolithic curves coarsely approximate different trends in monolithic, two-qubit gate infidelity as devices scale past the size of the chiplet (27 qubits).}
\label{fig:graph-fid-ecc}
\end{figure}

We estimate SWAP chain process fidelity using $F_{CX, chip}=1-0.0045$ based on the hybrid surface and Bacon-Shor error threshold~\cite{chamberland2020topological}. As a note, the infidelity of the hybrid Bacon-Shor threshold is half of the observed average CX infidelity of \textit{ibmq\_mumbai}, $1-F_{CX,Mumbai}$. All other parameters are the same as described in Section~\ref{sub:est-fid-near}. Figure~\ref{fig:graph-fid-ecc} presents SWAP chain process fidelity in a system where the on-chip qubits demonstrate average CX fidelity required for hybrid Bacon-Shor. However, since $F_{link}$ ranges from 0.86-0.96, the described chiplet systems are still not robust enough to implement error correction - either link quality must improve, or even better error rates in the local processor may compensate. Figure~\ref{fig:graph-fid-ecc} includes a comparison of total process fidelity of qubit state transfer on the 27-qubit chiplet architecture  vs. monolithic. Fidelity values are noticeably higher than those featured in~\ref{fig:graph-fid}, but the qualitative relationships between ranges chiplet vs. monolithic advantage remain.

All but two of the monolithic SWAP chain process fidelity curves in Figures~\ref{fig:graph-fid}(a-b) eventually intersect with the upper bound chiplet process fidelity, regardless of base chip size. If a negative correlation between qubit count and gate error cannot be avoided, chiplet architectures may provide the best route to boosting QC frontiers.

%We also consider directly analyzing the \dfour model. For the distributed heavy-hex system shown in Figure \ref{fig:topology-compare}, we may simplify the calculation by noting that a path from the top to bottom or from left to right passes through 10 qubits, requiring nine local SWAPs. We set $m=9$ in Equation \eqref{eq:dddrcompose}. This is not exactly the longest path through the device, but it's within a few SWAPs per chip. We ignore this approximation error here. For compatibility with Figure \ref{fig:graph-fid}, we use the ratio between reported CX error rates in these devices to suggest a ratio of 0.43 of the \textit{ibmq\_mumbai}'s \dfour parameters to those of the \textit{ibmq\_montreal}, scaling parameters to roughly reproduce the fidelity scaling of Equation \eqref{eq:mono-calc}. These results appear as Figure \ref{fig:chanl-fid}.
%\begin{figure}[h!]
%\includegraphics[width=0.49\textwidth]{figures/mono_chiplet_compare.png}
%\caption{Fidelity vs. system size for monolithic and chiplet architectures as calculated via the \dfour model in Equation \eqref{eq:dddrcompose}. Compares roughly with Figure \ref{fig:graph-fid}.}
%\label{fig:chanl-fid}
%\end{figure}
%Figure \ref{fig:chanl-fid} shows qualitatively similar results to the fidelity-based calculations in Subsection \ref{sub:est-fid}.

\begin{figure}[h!]
\centering
\includegraphics[width=0.49\textwidth]{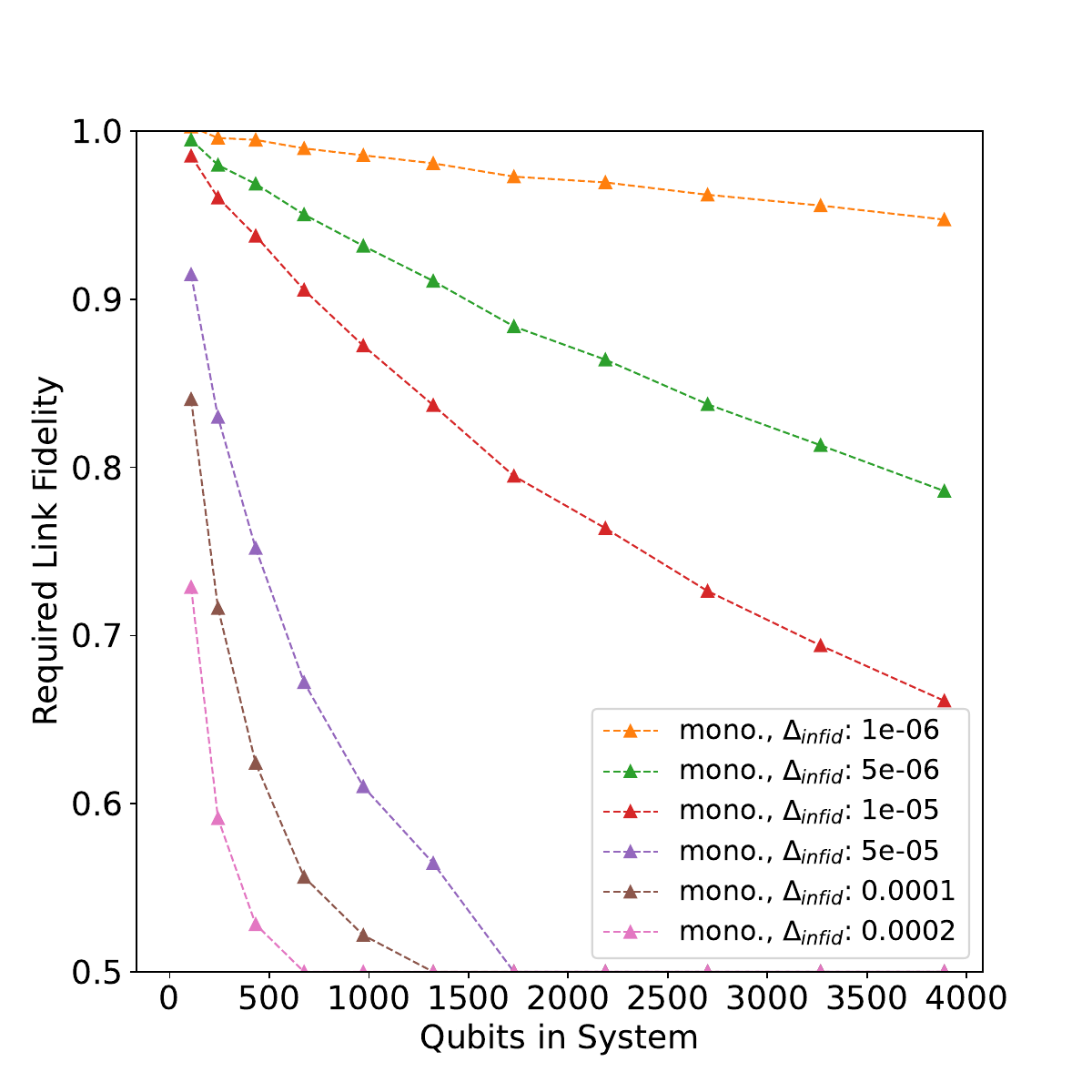}
\caption{Required link fidelity vs. qubits for equivalent diameter SWAP chain process fidelity between monolithic and 27-q chiplet architecture. The curves here use the same projected rates of change for the negative correlation between monolithic system size and average two-qubit gate fidelity as seen in Fig~\ref{fig:graph-fid}. In this model, $F_{CX,chip}=1-0.0045$.  As $\Delta_{infid.}$ increases, link fidelity thresholds in competing chiplet architectures can be fairly low.}
\label{fig:link-fid-vs-qubits}
\end{figure}

We expand our modeling to gain a better understanding of the link fidelity needed to achieve equal network diameter SWAP chain process fidelity for the chiplet and monolithic architectures of size N. We set our sights on fault tolerance, so in this analysis, we utilize on-chip thresholds required for hybrid surface and Bacon-Shor codes.  Figures~\ref{fig:link-fid-vs-qubits}(a) and (b) include a plot of required link fidelity, $F_{link}$ vs. qubits for the 27-qubit chiplet architecture, respectively, when compared to corresponding monolithic implementations. The curves here use the same projected rates of change for the negative correlation between monolithic system size and average two-qubit gate fidelity as seen in Fig~\ref{fig:graph-fid}. In this model, $F_{CX,chip}=1-0.0045$.  As $\Delta_{infid.}$ increases, link fidelity thresholds required for competing chiplet architectures to match monolithic SWAP chain process fidelity are significantly reduced in systems up to thousands of qubits in size.

Simple, 2-D, distributed architectures may show advantages over analogous, monolithic architectures even with modest improvements in link transfer fidelity. The point at which a chiplet architecture becomes favorable depends strongly on the relationship between monolithic chip size and noise that influences average two-qubit gate fidelity. Further experiments in this area will illuminate and clarify these tradeoffs, so we include a spread of possible values.

Though Figure \ref{fig:link-fid-vs-qubits}(a) and (b) use a fixed local error rate for the 27-qubit chiplet baseline, we show here that it is invariant under particular re-scalings of $R_{CX}$, $R_{link}$ and $\Delta$. Let $T_{link}$ be the total number of link uses to cross the graph diameter of a chiplet architecture and $T_{SWAP}$ be the total number of local SWAP gates. Assuming the chiplet architectures of Figures \ref{fig:topology-compare}(b) and (c), $T_{SW} + T_{link}$ is the total number of qubit movements in an architecture. Let $R^{(tot)}_{chip}$ be the multiplicative total for the chiplet architecture, $R^{(tot)}_{mono}$ that for the monolithic. Then
\begin{equation}
\frac{R^{(tot)}_{chip}}{R^{(tot)}_{mono}}
	= \frac{R_{link}^{T_{link}} R_{CX}^{3 T_{SWAP}}}{(R_{CX} + (N - n_{chip}) \Delta_{infid.})^{3 (T_{SWAP} + T_{link})}}.
\end{equation}
By matching powers, we may simultaneously scale  $R_{SW}$ and $\Delta_{infid.}$ by a multiplicative factor $a > 0$ while scaling  $R_{link}$ by $a^3$ without changing the ratio. On one hand, this shows that a relatively small improvement in CX gate performance is equivalent to a larger factor in link performance, because 3 CX gates form a local SWAP. In contrast, this ratio would be equal between local SWAP and link performance, so that aspect is largely an arbitrary detail of the construction. More broadly, however, the invariance shows the ratio of $R_{SW}$ and $\Delta$ to $R_{link}$ determines the tradeoffs shown in Figure \ref{fig:link-fid-vs-qubits}, rather than the specific choice of $R_{CX}$. Hence variation in required $R_{link}$ effectively includes the possibility of expected improvements in local errors up to rescaling.

\subsection{Modeling fragments of computation}
Finally, we depart from graph-driven analyses and consider the fidelity of small computation fragments with or without a link. The network-based comparisons as in Subsection \ref{sub:est-fid} avoid this problem by analyzing only the movements of qubits as a building block of computation. That approach may however miss aspects of realistic computation, in which many operations occur simultaneously with qubit movements. Examining realistic computations seems more favorable toward chiplet architectures under the size-noise tradeoffs described in Subsection \ref{sub:est-fid}, because while the impact of link transfers should remain comparable, the effects of local noise will be amplified when there are many local gates happening alongside qubit movements.

\begin{figure}[h!]
\includegraphics[width=0.4\textwidth]{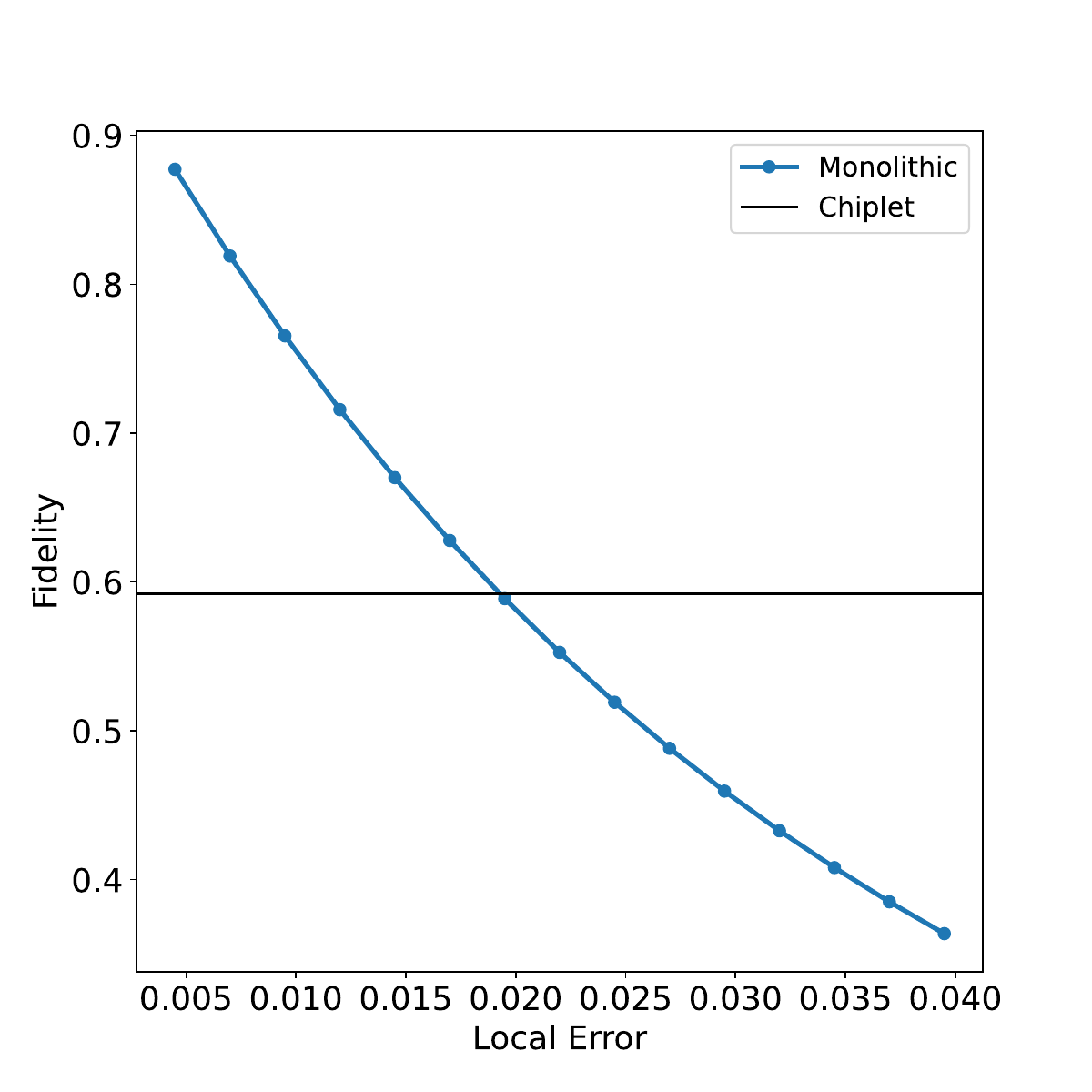}
\caption{Fidelity between simulated and ideal output state vs. monolithic error for 6-qubit random circuits. The black, horizontal line represents a simulated fidelity, while the decaying blue curve shows fidelity with increasing local error in a simulated monolithic configuration.}
	\label{fig:rcircs}
\end{figure}

Directly analyzing quantum circuits is limited, because the regimes in which distributed QC probably shows advantages are well beyond the capabilities of classical simulators. To get around the difficulty in simulating large quantum computations, we consider small, random circuits as models of computation fragments. Using Qiskit's built-in subroutine, \textit{qiskit.circuit.random.random\_circuit}, we generate 30 random circuits on 6 qubits. These are mapped to a custom, simulated, 6-qubit, bipartite backend. In the chiplet case, each 3-qubit cluster represents a fragment of a chiplet with internal all-to-all connections, while the middle connection models a link. This layout is inspired by that in \cite{zhong2021deterministic}. In the monolithic case, each of the three qubits connects on either side connects to one on the other. The backend is configured to assign depolarizing noise to local CX gates and amplitude-damping noise to the modeled link. We then compare the average fidelity with an ideal circuit for each of 30 noisy circuits with the ideal circuit.

Specifically, local CX gate noise is chosen to range from 0.0045 to 0.0395 in increments of 0.025 for the ``monolithic" case, extrapolating upward from the hybrid, Surface and Bacon-Shor threshold as considered in \ref{sub:est-fid-near}. Link noise is chosen to apply a 1\% depolarizing channel followed by an amplitude-damping channel with $\eta = 0.12$ based on channels modeled in Subsection \ref{sub:chanmod}. The results appear as Figure \ref{fig:rcircs} and show a qualitatively similar pattern to those of Subsection \ref{sub:est-fid}. The almost perfectly linear appearance of the plot is due to Qiskit's idealized density matrix comparison, which does not introduce measurement randomness as would tomography.

\subsection{An error detection scheme for links} \label{sec:errdetect}
While several existing studies propose error correction for amplitude-damping noise \cite{chuang1997bosonic,fletcher2008channel-adapted,shor2011high,grassl2014quantum,chessa2021quantum}, we focus on error detection. In the near term, small, noisy chiplets will probably benefit more from error detection and post-selection than from attempting to correct errors. In the long-term, larger, low-noise chips will probably use entanglement distribution followed by teleportation of data qubits, such that a detect-and-retry scheme is likely more efficient than using error correction around links.

Amplitude-damping errors are perfectly detected for any encoding into a subspace with a fixed number of computational basis of 0 and 1 states. Let $m$ be a number of encoded qubits. For any $m > k \in \NN$, there is a subspace of dimension $(m \text{ choose } k)$ spanned by states that include exactly $k$ qubits set to ``1" in the computational basis. Though this subspace is defined in the computational basis, it extends by linearity to a full Hilbert space of this dimension, supporting superpositions, entanglement, etc. Should an amplitude damping error occur, it will necessarily leave this subspace. Hence states in this subspace have perfect error detection.

We may therefore encode $\lfloor \log_2 (m \text{ choose } k) \rfloor$ qubits into $m$ qubits, where the probability of an error occuring is roughly proportional to $k$, and $\lfloor \cdot \rfloor$ denotes the floor function. We always obtain maximal encoding efficiency with $k = m/2$ but in some cases may optimize error probability by reducing the value of $k$. By Stirling's approximation and letting $\lceil \cdot \rceil$ denote the ceiling function, we find
\[ n \geq m - \lceil \frac{1}{2} \log_2 (\pi m ) \rceil \]
is achievable, implying asymptotically high efficiency upon success in return for a (detected) failure probability that scales with $k$.

\begin{figure}[t!]
\centering
\includegraphics[width=0.3\textwidth]{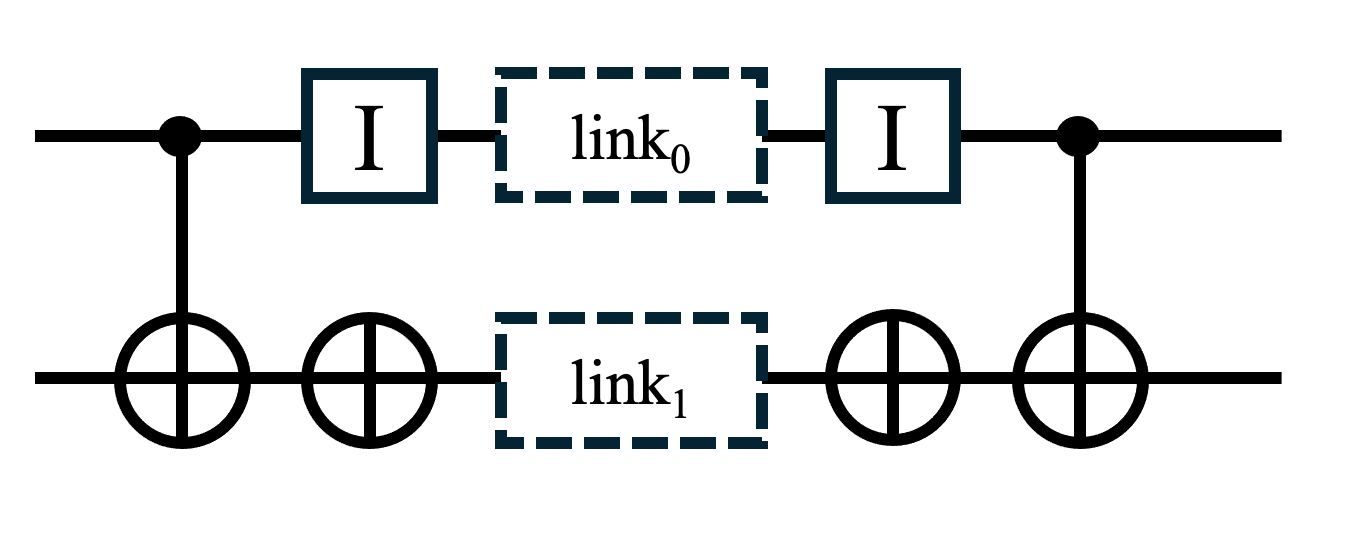}
\caption{Basic error detection for amplitude-damping links. The ``$+$" symbols represent $X$ and $CX$ gates, the ``$I$" a trivial identity gate, and the two dashed boxes are two copies of the link.}
\label{fig:err-corr}
\end{figure}

The simplest example of such a code is implemented by the transformation
\begin{equation}
\alpha \ket{0} + \beta \ket{1} \leftrightarrow \alpha \ket{10} + \beta \ket{01} ,
\end{equation}
a qubit version of the \cite[Equation 2.15]{chuang1997bosonic}. This is illustrated via Figure \ref{fig:err-corr} A slightly more sophisticated code on 2 qubits is given by
\begin{equation} \label{eq:twolinkcode}
\begin{split}
& \alpha_0 \ket{00} + \alpha_1 \ket{01} + \alpha_2 \ket{10} + \alpha_3 \ket{11} 
	\\ & \hspace{2mm} \leftrightarrow \alpha_0 \ket{0001} + \alpha_1 \ket{0010} + \alpha_2 \ket{0100} + \alpha_3 \ket{1000} ,
\end{split}
\end{equation}
which uses 4 qubits to encode 2 but still has $k = 1$. Table \ref{tab:ecc} summarizes encoding schemes for up to 10 qubits at a time.

\begin{table}
\centering
\begin{tabular}{| c | c | c | c |} \hline
Q Sent & Q Used & Err Mult & Efficiency \\  \hline
1 & 2 & 1 & 0.50 \\
2 & 4 & 1 & 0.50 \\
3 & 3 & 5 & 0.60 \\
4 & 6 & 3 & 0.67 \\
5 & 7 & 3 & 0.71 \\
6 & 8 & 4 & 0.75 \\
7 & 10 & 4 & 0.70 \\
8 & 11 & 4 & 0.73 \\
9 & 12 & 5 & 0.75 \\
10 & 13 & 5 & 0.77 \\ \hline
\end{tabular}
\caption{Table of encodings for up to 10 qubits. ``Q Sent" and ``Q Used" refer respectively to the number of qubits sent and the number used (including those for error detection). The error multiplier is the number of ``1"s in the computational basis, such that this number times $\eta$ is the total probability of a detected failure. The listed efficiency is the number of qubits sent divided by the number used, conditioned on successful transmission. \label{tab:ecc}}
\end{table}

Because detected errors are more likely with larger encodings, the overall probability of success for one use of the code falls with the number of qubits transmitted. In a fully post-selected application, encoding more qubits at once nonetheless has advantages both in overhead and total failure likelihood. In a detect-and-retry scheme, there might be advantages to distributing entanglement more piecewise, since each detected failure then has a lower cost in terms of lost resources.

With a CX error of about 0.01 depolarizing noise and a 0.02 readout error, we may roughly model the conditional channel applied on a ``success" measurement as probably about 0.03-0.04 depolarizing with current local error rates. The 0.88 reported efficiency from \cite{zhong2021deterministic} and 0.02 readout error yield a post-selection efficiency of about 0.86. As mentioned in that paper, however, it should be possible to increase the efficiency in future experiments. Similarly, \cite{magnard2020microwave} suggests a potential transfer fidelity of 0.96. Error detection with post-selection is potentially most useful in relatively small, early distributed experiments. Post-selection is not ideal for long SWAP chains with many potential failure points, but it may perform better in the tree or expander connectivities as proposed in Section \ref{sub:graph-diameter}. Logarithmic depth circuits with expanderized connectivity woud have inverse polynomial success probability even without retrying any transfers mid-computation.

Broadly, we expect on-chip noise to fall to extremely small values with continued engineering research. If the fidelity of link transfers plateaus at a lower value, such as the 0.96 mentioned in ~\cite{magnard2020microwave}, then comparisons such as those in Section \ref{sec:networklayer} would ultimately show a disadvantage to using links with extremely high fidelity of local processing. In this regime, error detection may enable continued advantages in chiplet architectures. With sufficiently good on-chip processing and readout, one may consider an abstracted transfer operation that distributes entanglement, detects errors until retrial, and ultimately teleports the computation qubits. Such a scheme may still have a smaller on-chip footprint than error correction, requiring only two qubits in principle. Hence link error detection could be most valuable in a regime of small, high-quality chiplets.

% As a simple example, we generate a 6-qubit, depth 3 random circuit using Qiskit's \textit{random\_circuit} method (shown in Figure \ref{fig:rcircs}), transpile it to a simple connectivity given by a pair of 3-qubit rows connected via the middle. We compute Hellinger fidelity with the noiseless case, measuring the similarity of measurement outcomes, when the middle connection is a link with depolarizing $r$ of 0.10 to simulate a CX gate and local noise of 0.01, finding a result of 0.74. When the link is instead a local connection with 0.01 CX noise, the Hellinger fidelity is much higher at 0.87. It however drops to a value of 0.75 with all-local gates having 0.02 CX noise, and to 0.66 with 0.03 CX noise. This test loosely hints at a tradeoff between links introducing errors and potentially improving local noise.

\section{Discussion and Conclusions}
The main result of this work is evidence for the role of distribution as a quantum computing paradigm.
Effective modeling requires choosing good abstractions. It would not be feasible to model a complete Hamiltonian when considering quantum algorithms, as the amount of detail would overwhelm any simulator and require specifying far too many parameters. In contrast, a model that abstracts away all physical reality might not usefully determine whether proposed applications can actually use proposed systems. While our analysis is largely focused on a snapshot in time during the NISQ era of quantum computing (2021), we anticipate that the models described here will remain relevant as link and on-chip fidelities improve in lockstep. An open-source artifact associated with the work presented in this paper can be found in Ref.~\cite{links-repo}.

A key outcome of this work has been to find essential aspects and constraints. Noise distinguishes links from local transfers, and depending on the network topology, bandwidth and distance may be important. In contrast, the latency of a single microwave link is small, and the success rate high enough to use without local distillation. These aspects make microwave-linked chiplets a promising architecture for scaling up arrays of physical qubits.

We have taken care to minimize idealized future technologies, so that our work may bridge understanding between different layers. As much as possible, we start from the experimentally implemented hardware and build up to the level of potential applications. Nonetheless, it is impossible to cover everything or be completely sure at each layer, because we continue to study systems that are not yet built. Here we highlight some of these open questions as promising for further study:
\begin{enumerate}
\item In Section \ref{sec:networklayer}, we compare possibilities for state-of-the-art microwave links to connect current, industrial-quality chips. A major experimental question is how and if these technologies will be combined. To account for possible impacts of integration and unpredictability of future developments, we consider a wide range of possible fidelities for both on-chip operations and links. An essential next step in distributed quantum computing is experimental work to begin resolving these ambiguities.
\item The analysis of Section \ref{sub:est-fid} hints that for plausible ranges of noise parameters, the overall error probability for simultaneous operations linking many qubits will be lower in a chiplet than in a monolithic architecture. Intuitively, a logical qubit spread over a chiplet system could have looser fault-tolerance requirements than one implemented monolithically. Quantum error correction is a major topic of its own with many complexities, options, and subtleties, so it is beyond the scope of this paper to rigorously investigate this intuition. Nonetheless, it seems likely that in contrast to common expectations, distribution may usefully lie below the logical qubit layer in a fault-tolerant scheme.
\item Today's optical and microwave links represent vastly different regimes in expected range and performance as described in Section \ref{sec:physicallayer}. There is nonetheless optimism that in the long term, matter-optical transduction will benefit from major advances in hardware and local error reduction. It remains open if local networks could eventually use optical fiber to similar effects as proposed for microwave links in this work.
\item Another recent work \cite{zhou2023realizing} proposes a quantum router, using more specialized quantum hardware to achieve all-to-all connectivity dynamically. One might expect that such a router, like a hierarchical graph, achieves high sequential connectivity and low graph diameter but may suffer bottlenecks at the router. Nonetheless, dynamical switching might improve connectivity while better isolating qubits against unintended interactions.
\item As quantum computers increase in size, efficient algorithms to optimize the flow of data around a device increase in importance. The development of these compilation and mapping techniques will enable analyses to better quantify the costs of different connectivities.
\end{enumerate}
Distributed quantum computing encompasses a wide range of possibilities. While much research focuses on applications of long-range networks of logical quantum computers to communication problems, we emphasize ways in which short-range, physical networks could accelerate the path to fundamental abstractions and protocols.

\section{Acknowledgments}
NL and KS were IBM Postdocs at UChicago and the Chicago Quantum Exchange.  Frederic T. Chong is Chief Scientist at Super.tech and an advisor to Quantum Circuits, Inc. This work is funded in part by EPiQC, an NSF Expedition in Computing, under grants CCF-1730082/1730449; in part by STAQ under grant NSF Phy-1818914; in part by DOE grants DE-SC0021526, DE-SC0020289, and DE-SC0020331; in part by NSF OMA-2016136 and the Q-NEXT DOE NQI Center; and in part by NSF grant OMA-1936118. This research used resources of the Oak Ridge Leadership Computing Facility at the Oak Ridge National Laboratory, which is supported by the Office of Science of the U.S. Department of Energy under Contract No. DE-AC05-00OR22725.

We thank Paul Magnard, David Schuster, Andrew Cross, Oliver Dial, and John Smolin for helpful feedback during the research and writing of this paper.

\bibliographystyle{quantum}
\bibliography{refs}

\end{document}